\documentclass[%
 reprint,
 amsmath,amssymb,
 aps,nofootinbib,   
]{revtex4-1}
\usepackage{bm}
\PassOptionsToPackage{linktocpage}{hyperref}
\usepackage[hyperindex,breaklinks]{hyperref}
\usepackage{enumitem}
\usepackage{slashed}

\renewcommand{\theta}{\vartheta}

\newcommand{\bra}[1]{\ensuremath{\left< #1\,\right|}}
\newcommand{\ket}[1]{\ensuremath{\left|\, #1\right>}}

\usepackage{array}
\usepackage{mathtools}

\usepackage{etoolbox}

\begin{document}

\title{Quantum Gravity in Species Regime}

\author{Gia Dvali  
} 
\affiliation{%
Arnold Sommerfeld Center, Ludwig-Maximilians-University,  Munich, Germany, 
}%
 \affiliation{%
Max-Planck-Institute for Physics, Munich, Germany
}%

\date{\today}

\begin{abstract} 

A large number of particle species
allows to  formulate quantum gravity in a 
special double-scaling limit, the species limit. 
In this regime,  quantum gravitational amplitudes simplify 
 substantially.  An infinite set of perturbative corrections, 
 that usually blur the picture,  vanishes, whereas 
 the collective and non-perturbative effects can be cleanly extracted.  
 Such are the effects that control physics of 
 black holes and of de Sitter and their entanglement curves.   
In string theory example, we show that the entropy of open strings 
matches the Gibbons-Hawking entropy of a would-be de Sitter state
at the point of saturation of the species bound.
This shows, from yet another angle, why quantum gravity/string theory cannot tolerate a de Sitter vacuum. Finally, we discuss various observational implications. 

  \end{abstract}

\maketitle

\section{Introduction and Summary} 
 
  The main technical obstacle in understanding the microscopic physics 
 of objects such as a black hole or a de Sitter space, 
  is a presumed complexity of quantum gravitational amplitudes  
  in non-perturbative regimes.  
  It is a common perception that processes associated 
  with non-perturbative objects, at weak coupling 
  $\alpha$, are exponentially suppressed 
  ${\rm e}^{-1/\alpha}$. \\
  
  From here, one is tempted to makes two conclusions. 
 First, non-perturbative effects are completely dwarfed  
  by a variety of contributions that are power-law in $\alpha$.
Secondly, at weak coupling, the non-perturbative objects, for all 
practical purposes, can be treated 
semi-classically.  \\

  However, in several instances the situation is much more subtle.
This is the case for the states such as 
 a black hole or a de Sitter \cite{Dvali:2011aa, Dvali:2012en, Dvali:2013eja, Dvali:2014gua, Dvali:2017eba}. 
   The reason is that both states belong to the category of 
 the saturated systems, or {\it saturons} for short \cite{Dvali:2020wqi}. 
Such systems, despite being non-perturbative and macroscopic, depart from semi-classicality rather fast.   
 \\ 
    
  For understanding the meaning of saturation, let us first note that 
non-perturbative processes can be  
  strongly enhanced by the following three factors: 
  \begin{itemize}
  \item The occupation number of quanta, $N$; 
  \item The micro-state entropy, $S$; 
  \item  The number of interacting particle species, $N_{\rm sp}$. 
\end{itemize}
  In fact, the enhancement can be so strong that a given process can be 
  pushed to saturating the unitarity bound.  
 The phenomenon is not limited to gravity and is a generic  property 
  of {\it saturons}. 
  
\subsection{Saturons}     
    
 This term describes the systems that satisfy the following limiting relation, imposed by unitarity, \cite{Dvali:2020wqi}:  
     \begin{equation}\label{saturation}  
   N = \frac{1}{\alpha} = S \,.
  \end{equation} 
  Here, $\alpha$ must be understood as the scale-dependent quantum coupling of the theory,  evaluated at the size of the system, $R$. 
 As usual, the entropy $S$ is defined as the log from the number of 
 degenerate micro-states.   
   \\

The saturation condition (\ref{saturation}) encodes three relations. 
  The primary idea \cite{Dvali:2020wqi, Dvali:2019jjw} is that, for an arbitrary system at weak coupling,
unitarity imposes the following upper 
bound on the entropy,
   \begin{equation}\label{S111}  
   S \leqslant \frac{1}{\alpha} \,.
  \end{equation} 
 Correspondingly, a self-sustained system with $N$ constituent quanta, 
 at its maximal entropy capacity, satisfies the relation (\ref{saturation}). \\

 It was also argued in \cite{Dvali:2020wqi, Dvali:2019jjw}
that for a self-sustained system, satisfying (\ref{saturation}), 
the entropy $S$ is equal to the area measured in units  
of a Goldstone decay constant, $f$. 
The presence of a Goldstone is generic,
due to a spontaneous breaking of Poincare symmetry by any saturon.
Thus, the saturation relation (\ref{saturation}), also implies the 
area-form of the entropy\footnote{Throughout the paper, the unimportant numerical coefficients shall be set equal to one.},   
     \begin{equation}\label{SA}  
   S = {\rm Area} \cdot f^2 = (Rf)^2 \,.
  \end{equation}  
  For definiteness, we wrote the expression in four space-time dimensions, but the same area-law holds in arbitrary
  dimensions.  \\

Notice, in case of a black hole, the Goldstone of 
a spontaneously broken Poincare symmetry, is the graviton mode.
Its decay constant is a Planck mass $f=M_P$. This, was argued in 
 \cite{Dvali:2020wqi, Dvali:2019jjw}, shows that the area-form of the 
 Bekenstein-Hawking entropy \cite{Bekenstein:1973ur}, 
     \begin{equation}\label{BH}  
   S = (RM_P)^2 \,, 
  \end{equation} 
is a particular example of the area-form of the entropy 
(\ref{SA}), originating from the phenomenon of saturation.
The same is true about Gibbons-Hawking entropy \cite{Gibbons:1977mu} of de Sitter.  This entropy is also given by 
(\ref{BH}), with $R$ understood as the de Sitter curvature radius. 
  \\

That is, both, a black hole and a de Sitter Hubble patch, are saturons, 
as they satisfy (\ref{saturation}) and (\ref{SA}). 
 However, the above correlations appear to be universal and to go well beyond gravity \cite{Dvali:2020wqi, Dvali:2019jjw}.  The meaning of the expressions (\ref{saturation}) and (\ref{SA}), shall become more transparent later.  \\

 The goal of the present paper is to focus on the role of the third item, 
 the number of particle species $N_{\rm sp}$. 
 Throughout the paper, unless otherwise stated, 
 species will be assumed to be lighter that the relevant energy 
 scale in a given process.  We shall primarily be interested in the species of low spin, although most of the results will be applicable in the 
 presence of spin-$2$ species, as it is the case in Kaluza-Klein theories. \\

 The idea is to use the magnifying power 
 of particle species \cite{Dvali:2007hz}, for cleanly extracting certain quantum gravitational 
 effects. We then apply this method to gravitational saturons in form 
 of black holes and de Sitter.  \\

 \subsection{Species Limit} 
 
  We shall define a double-scaling
 limit in which the Planck mass $M_P$ and the number of species $N_{\rm sp}$ are taken infinite while 
  the following relation among them is maintained: 
        \begin{equation}\label{SP1}  
   M_P \rightarrow \infty,~   N_{\rm sp}  \rightarrow \infty\,, 
   ~ \frac{M_P}{\sqrt{N_{\rm sp}}} = {\rm finite} \,.
  \end{equation} 
  This regime has to be supplemented by a proper scaling 
  of the graviton occupation number $N$. 
 We shall refer to (\ref{SP1}), as the {\it species 
 limit}. \\
 
 The significance of the species limit (\ref{SP1}), from quantum gravitational
 perspective, is immediately captured by the fact that in this limit 
 the half-decay time of a black hole, 
       \begin{equation}\label{H111}  
   t_{\rm half} = R^3 \frac{M_P^2}{N_{\rm sp}} \,,
  \end{equation} 
is finite, despite the fact that the black hole mass is infinite. 
The microscopic meaning of this phenomenon, shall be discussed in details.  
\\

 In the species regime (\ref{SP1}), the quantum gravitational amplitudes simplify
 significantly. The secondary processes 
 vanish while the non-perturbative effects of interest stay finite. 
 In other words,  the species limit (\ref{SP1}), allows to distill a set of 
essential quantum gravitational effects. 
Some aspects of this extraction were already studied in \cite{Dvali:2020wqi, Dvali:2020etd}, on which we shall expand.   
\\

  Next, we shall apply this treatment to saturons in gravity. 
  As already said, such are black holes and de Sitter. The fact that these states represent 
saturons, follows from 
 the $S$-matrix formulation of 
quantum gravity/string theory. 
 In this formulation, both objects must be considered as 
 excited composite states constructed on top of a valid
 $S$-matrix vacuum, such as Minkowski.  
 In other words, the $S$-matrix formulation leads to their {\it corpuscular 
 resolution}. \\

 The corpuscular picture 
 \cite{Dvali:2011aa, Dvali:2012en, Dvali:2013eja},  
 reveals that black holes as well as de Sitter satisfy the saturation relation 
 (\ref{saturation}).  
  Due to saturation, despite of their non-perturbative 
  nature, both systems are susceptible to quantum gravitational 
  effects that alter their semi-classical properties on rather  
  short time-scales. The relevant  time-scales are only 
  power-law (or even logarithmic \cite{Dvali:2013vxa}) in $1/\alpha$. 
  
  \subsection{Inner Entanglement and Quantum Break-Time}

 A particularly important time-scale, delivered by the corpuscular 
 picture, is the {\it quantum break-time} $t_Q$.
  The concept 
 was introduced in 
 \cite{Dvali:2013vxa}, within a prototype model for a black hole
 $N$-portrait \cite{Dvali:2011aa, Dvali:2012en}.  It was applied to de Sitter and to black holes 
 in \cite{Dvali:2013eja} and in subsequent papers. 
 The physical meaning of $t_Q$ is that, beyond it, a total breakdown of the semi-classical approximation
 takes place. The major mechanisms contributing into this departure 
 are: 1) {\it Inner entanglement}
 \cite{Dvali:2013eja, Dvali:2014gua, Dvali:2017eba}; and 2) {\it Memory burden} effect \cite{Dvali:2018xpy, Dvali:2018ytn, Dvali:2020wft}.  
 Both effects shall be reviewed in due course.  \\

As discussed originally in  \cite{Dvali:2013eja},  the above effect is fundamental for understanding physics behind the so-called Page's curve \cite{Page:1993wv}.  According to Page, 
the entanglement must reach the maximum after a black hole emits  
about half of its mass.  The semi-classical picture cannot explain this.
This picture is blind to an inner structure of the black hole.
 Because of this, the semi-classical theory captures only one 
sort of entanglement: The entanglement between a black hole and an outgoing radiation.  
 \\

The corpuscular picture reveals the existence of
another type of entanglement, which we can refer to as  the {\it inner entanglement}  \cite{Dvali:2013eja}. This effect describes  the entanglement among the internal degrees of freedom composing a black hole.  
Together with the memory burden effect
\cite{Dvali:2018xpy, Dvali:2018ytn, Dvali:2020wft}, the inner entanglement grows in time  and reaches the maximum, latest,
by a half-decay. This is the origin of the quantum break-time $t_Q$. 
  These effects offer a new physical meaning for the 
Page's curve. \\

 Note, by providing an internal quantum clock, the corpuscular 
picture predicts the existence of the entanglement curve also for de Sitter \cite{Dvali:2013eja}.  It shows that the entanglement must reach its maximum at the quantum breaking point. In this sense, de Sitter is similar to a black hole. 
However, beyond $t_Q$, the  
time evolutions of a black hole and of de Sitter exhibit 
drastic differences.  While a black hole can continue its existence 
beyond $t_Q$, for de Sitter this is deadly. 
  The incompatibility of a de Sitter vacuum  with the $S$-matrix formulation 
 of quantum gravity, manifests itself though the anomalous quantum 
  break-time \cite{Dvali:2020etd}. \\

  In the present paper, we shall re-analyse the above 
  dynamics, by taking into account the effect of species. 
  It has already been shown previously that, with other 
 parameters fixed, the species shorten 
  $t_Q$.  More precisely, for a generic saturated system, $t_Q$ scales as 
  \cite{Dvali:2017eba, Dvali:2020wqi, Dvali:2020etd},
      \begin{equation}\label{QQQ}  
  t_Q \, = \,  \frac{R}{\alpha N_{\rm sp}} \,.
  \end{equation} 
 Notice, for a black hole or a de Sitter, this time-scale is equal to (\ref{H111}).  \\
 
In the present case, we shall study the species effect in various double-scaling regimes, such as
 (\ref{SP1}).  This allows us to isolate the three important time scales of: 1) Half-decay; 2) Inner entanglement; 3) Memory burden.    All three time-scales are bounded by $t_Q$ (\ref{QQQ}) (and by $t_{\rm half}$
 (\ref{H111})) and remain finite in the species limit (\ref{SP1}). 
    \\
  
 \subsection{String Theory}

 Next, we use the species limit  (\ref{SP1}) 
 for monitoring 
 the resistance of string theory against the deformation towards 
 a de Sitter state. 
 We take this limit in an explicit string theoretic
 example 
 in which $n$ $D-\bar{D}$-brane pairs are plied up on top of each other.   
 In the species limit, the string coupling $g_s$ vanishes while 
 the number $N_{\rm sp} =n^2$ of Chan-Paton factors is taken infinite.
 At the same time, the product $ng_s$ is kept finite. 
 As already shown in \cite{Dvali:2020etd}, 
 this product determines the quantum break-time 
 $t_Q$ of a would-be de Sitter 
 state.  Using $ng_s$ as a control parameter, we can change 
 $t_Q$.  
 \\
 
 This example reveals some interesting correlations.  
 The entropy of open string modes, matches the Gibbons-Hawking 
 entropy of de Sitter space,  when the number of Chan-Paton species saturates the unitarity bound (\ref{S111}). 
 This bound, in the present case reads as, $ng_s \lesssim 1$.   \\
 
 At the same point, the thermal corrections from Gibbons-Hawking excitations of the open strings
 can flip the sign of the tachyon mass and stabilize it. 
 However, for the same values of the parameters, the de Sitter curvature approaches the 
 string scale.  Simultaneously, the quantum break-time becomes 
 of order the string length.
 This is the way the string theory responds 
 to its deformation towards a would-be de Sitter state.  \\
 
 We see that string theory possesses the resources, in form of 
 open string modes,
for accounting for Gibbons-Hawking 
 entropy. However, the very same degrees of freedom,  
speed-up the quantum breaking of the de Sitter ``vacuum". 
In this way, the string theory exhibits a built-in mechanism 
restoring the $S$-matrix consistency by abolishing de Sitter.  
 \\
  
   Finally, we discuss some observational imprints of 
   quantum gravity coming from inflation and 
 from  black holes.  By consistency, the duration of inflation 
 is bounded from above by $t_Q$. Since species
 shorten $t_Q$, they strengthen the quantum gravitational imprints 
 from the inflationary phase, making them potentially-observable, or 
 possibly dominant \cite{Dvali:2020etd}. \\
 
  Witten has introduced the concept of {\it meta-observables}  in 
  de Sitter \cite{Witten:2001kn}.  This concept is linked with the classical no-hair properties of de Sitter.  However, the finiteness of $t_Q$, effectively 
 can promote them into quantum-observables.  Basically, we are saying that $t_Q$ endows 
 de Sitter with a quantum hair \cite{Dvali:2013eja, Dvali:2018ytn}.   This hair 
 is suppressed by $1/t_Q$ and so are its observable imprints. 
  Since the species shorten $t_Q$ (see, (\ref{QQQ})), these imprints can be significant, even 
for non-far-future observations, 
if the number of species is high \cite{Dvali:2020etd}.  
This magnifying power, is demonstrated by the species limit 
(\ref{SP1}). In this limit,  the de Sitter quantum hair is non-vanishing, even though 
the Planck mass is infinite.

 \section{Species and Scale of Quantum Gravity} \label{SP}
 
 The fundamental property of species is the enhancement of 
the quantum gravitational effects.  
 This phenomenon is linked with lowering the gravitational cutoff, $M_*$, by species. 
 Here, the cutoff $M_*$ is defined as the scale at which 
 the quantum gravitational interaction becomes strong.
 In particular, a particle scattering process with momentum-transfer exceeding 
$M_*$,  cannot be described within the low energy QFT of 
gravity.  
Correspondingly, no semi-classical treatment is possible
for such systems.  In particular, this concerns the would-be 
cosmological scenarios with either temperatures or curvatures exceeding 
the scale $M_*$. \\

The scale $M_*$ is highly sensitive to the number of species 
$N_{\rm sp}$.  This follows from the arguments that 
are independent of perturbation theory. 
  In  particular, the black hole physics tells us \cite{Dvali:2007hz} that in theory with 
 $N_{\rm sp}$ particle species in $4$-dimensions, there is 
 the following upper bound on the scale
 $M_*$, 
   \begin{equation} \label{Mstar} 
  M_*  =   \frac{M_P}{\sqrt{N_{\rm sp}}} \,.
    \end{equation}   
   Correspondingly, in $d$ space-time dimensions the bound reads, 
   \begin{equation} \label{MstarD} 
  M_*  =   \frac{M_d}{N_{\rm sp}^{\frac{1}{d-2}}} \,, 
    \end{equation}       
    where $M_d$ is a $d$-dimensional Planck mass. 
   This bound is non-perturbative and cannot be removed by 
   re-summation.   
  One can arrive to it in several alternative ways. 
  Correspondingly,  the bound (\ref{Mstar}) can be given the 
 several different physical meanings. \\
 
  We shall briefly recount two physical 
arguments of \cite{Dvali:2007hz}.  More non-perturbative evidence supporting (\ref{Mstar}), shall emerge later.
 The first argument is based on the black hole evaporation. 
Within the validity of semi-classical gravity, a black hole
of mass $M$ and radius $R = M/M_P^2$, emits a thermal Hawking radiation
of temperature $T = 1/R = M_P^2/M$ \cite{Hawking:1974sw}.
 Because of its thermal nature  
and the universality of gravitational interaction, 
Hawking radiation is democratic in 
all particle species. 
 Due to this, the mass of a black hole changes in time according to the 
 following Stefan-Boltzmann law, 
\begin{equation}\label{SB}
    \dot{M} = - N_{\rm sp} T^2 \,. 
\end{equation}      
 Using the relation between the mass and the temperature, we can rewrite this equation in the following form, 
 \begin{equation}\label{nonT}
    \frac{\dot{T}}{T^2} =  N_{\rm sp} \frac{T^2}{M_P^2} \,. 
\end{equation}      
 This form is highly instructive. The left hand side of the equation 
 represents a measure of a departure from semi-classicality 
 and thermality. Within the validity of the semi-classical treatment, 
 this quantity must be less than one.  This puts an universal  
 upper bound on a temperature of a semi-classical black hole,  
 \begin{equation} \label{Tstar} 
  T_*  =   \frac{M_P}{\sqrt{N_{\rm sp}}} \,.  
    \end{equation}  
    This maximal temperature marks the gravitational cutoff of the 
low energy theory (\ref{Mstar}). An extrapolation of the 
semi-classical regime above this scale, would lead us into an obvious  
inconsistency. In particular, a black hole of 
temperature $T\gg T_*$, would lose energy faster than its own inverse temperature.   Notice, it would radiate its entire mass within
the time $\Delta t <  1/T \sim R$. This is absurd.
Under no circumstances can such a black hole be 
 treated semi-classically. 
\\

Alternatively, we can arrive to the bound (\ref{Mstar}), by considering a maximal 
possible Gibbons-Hawking 
temperature of a de Sitter like state in a theory with  
$N_{\rm sp}$.   
Since the pioneering work by Gibbons and Hawking \cite{Gibbons:1977mu}, it is well-known 
that a de Sitter like Universe (such as an inflationary Universe) is constantly creating particles with 
 thermal-like spectrum of temperature $T_{GH} = H$. 
 Here $H$ is the Hubble parameter, which is related with the de Sitter curvature (Hubble) radius as $H= 1/R$.  
 The value of 
 $H$, is determined by the energy density $V$ as, $H^2 = V/M_P^2$.   In inflation, $V$ is dominated by the potential energy density of the inflaton field. \\
 
Similarly to the Hawking radiation of a black hole, 
within the validity of semi-classical gravity, 
the Gibbons-Hawking radiation is nearly-thermal. 
Its energy density is given by, 
 \begin{equation} \label{Maxdensity} 
  \rho_{GH}  =   N_{\rm sp} T_{GH}^4 \,.  
    \end{equation}  
  It is easy to check that, for a temperature $T_{GH}  \gg M_*$, this expression would exceed the energy density $V$ of the de Sitter. Of course, this is impossible. \\
 
 For example, just like in the case of a black hole, 
 a Hubble patch with $H > M_*$, would convert its entire energy into the Gibbons-Hawking radiation in less than one Hubble-time, $\Delta t = 1/H$.  
 This is again an absurd. \\
 
 Thus,  similarly to the temperature of a black hole (\ref{Tstar}), 
 $T_{GH}$ is bounded by the scale  
 $M_*$ (\ref{Mstar}).  
  Equivalently, the inflationary Hubble scale is bounded as,  
         \begin{equation} \label{BoundH} 
  H \leqslant \frac{M_P}{\sqrt{N_{\rm sp}}} =M_* \,.
    \end{equation}  
 We shall adopt this as an absolute upper bound.
 Discussions of other aspects of species bound in de Sitter can be found in \cite{Dvali:2008sy}. 
   \\
 
  The bound (\ref{Mstar}) has also been derived 
  \cite{Dvali:2008ec, Dvali:2008jb, Brustein:2009ex, Palti:2019pca}
  from the requirement 
  that a black hole should not exceed 
  the maximal capacity of information storage and processing.  In particular, the species entropy should not 
  exceed the standard Bekenstein-Hawking entropy \cite{Bekenstein:1973ur} of a black hole.   
  \\

  In fact, the information storage capacity is also bounded by the unitarity of the scattering process in which two gravitons, 
of center of mass energy $E = N_{\rm sp} M_*$, produce
 $N_{\rm sp}$ species \cite{Dvali:2020wqi}.  The unitarity of this process reinforces the bound 
 (\ref{S111}), on the entropy of the final state.  This effectively translates as the bound (\ref{Mstar}).   
 This connection shall be discussed in more details below. \\
 
 There also exist perturbative arguments  supporting 
 (\ref{Mstar})  (see, e.g., \cite{Dvali:2001gx, Veneziano:2001ah}).
  We shall limit ourselves with non-perturbative ones, as 
 they cannot be avoided by the re-summation  of perturbation series.

 \section{Species Regime of Quantum Gravity}

  We have seen that  
   non-perturbative arguments unambiguously 
 indicate that quantum gravity is becoming strong 
 at the scale $M_*$, given by (\ref{Mstar}). 
 Already this fact shows that species possess a magnifying 
 power over quantum gravity.  Using this power, 
 we would like to cleanly extract certain quantum gravitational effects. \\
   
For this, we first define a {\it basic}  quantum 
gravitational coupling $\alpha$.  Let us consider a tree-level 
scattering process involving gravitons with momentum-transfer
$q$. For example, an annihilation of a pair of gravitons, of a 
center of mass energy $\sim q$, into a pair of some species. 
This process is governed by an effective quantum coupling 
given by 
 \begin{equation} \label{Alpha}  
\alpha =  \frac{q^2}{M_P^2}  \,.   
 \end{equation}
 This is what we shall call a basic quantum gravitational coupling. 
 In general, in a process with more complicated diagrammatic structure, 
 the coupling $\alpha$ in each vertex is determined by 
 the momentum flow through it. 
 \\
 
The momentum-transfer in a graviton vertex 
shall be restricted by the cutoff $q \leqslant M_*$. Correspondingly, 
 the coupling shall satisfy 
 \begin{equation} \label{AlphaM}  
\alpha \leqslant  \frac{M_*^2}{M_P^2} = \frac{1}{N_{\rm sp}} \,.   
 \end{equation}
In what follows, we shall work in the regime of a weak 
coupling $\alpha$ and effectively take the limit, 
 \begin{equation} \label{AlphaL}  
\alpha \rightarrow 0 \,. 
 \end{equation}
Naively, one may expect that, in this limit, all quantum gravitational effects 
vanish. However, this is not the case. 
In reality, (\ref{AlphaL}) allows to make certain phenomena more transparent.  For achieving this, the scalings of 
other quantities must be chosen properly. For the convenience of the description,  we shall introduce some useful parameters.

\subsection{Species Limit} \label{SLimit}  

 First, let us  define a {\it species coupling}, 
  \begin{equation} \label{Lambda}  
\lambda \equiv  N_{\rm sp} \alpha = 
 \frac{q^2}{M_*^2} \,,  
 \end{equation}
where in the very last expression we used the 
relations (\ref{Mstar}) and (\ref{Alpha}).   \\

 For a reader familiar with the planar treatment 
 of QCD with large number of colors \cite{tHooft:1973alw}, 
it must be clear that  the species coupling (\ref{Lambda}) 
represents a gravitational  
 analog of 't Hooft's coupling.  
 Due to universality of the gravitational interaction, the  role of ``colors" in gravity is played by the number 
 of all particle species. \\
 
 This said, there are fundamental differences. Unlike the gluons in QCD, which carry color and anti-color indexes, no species label is carried by the graviton.  This suppresses a large number of quantum 
 gravitational processes analogs of which in gluodynamics would be non-zero.  \\

\begin{figure}
 	\begin{center}
        \includegraphics[width=0.53\textwidth]{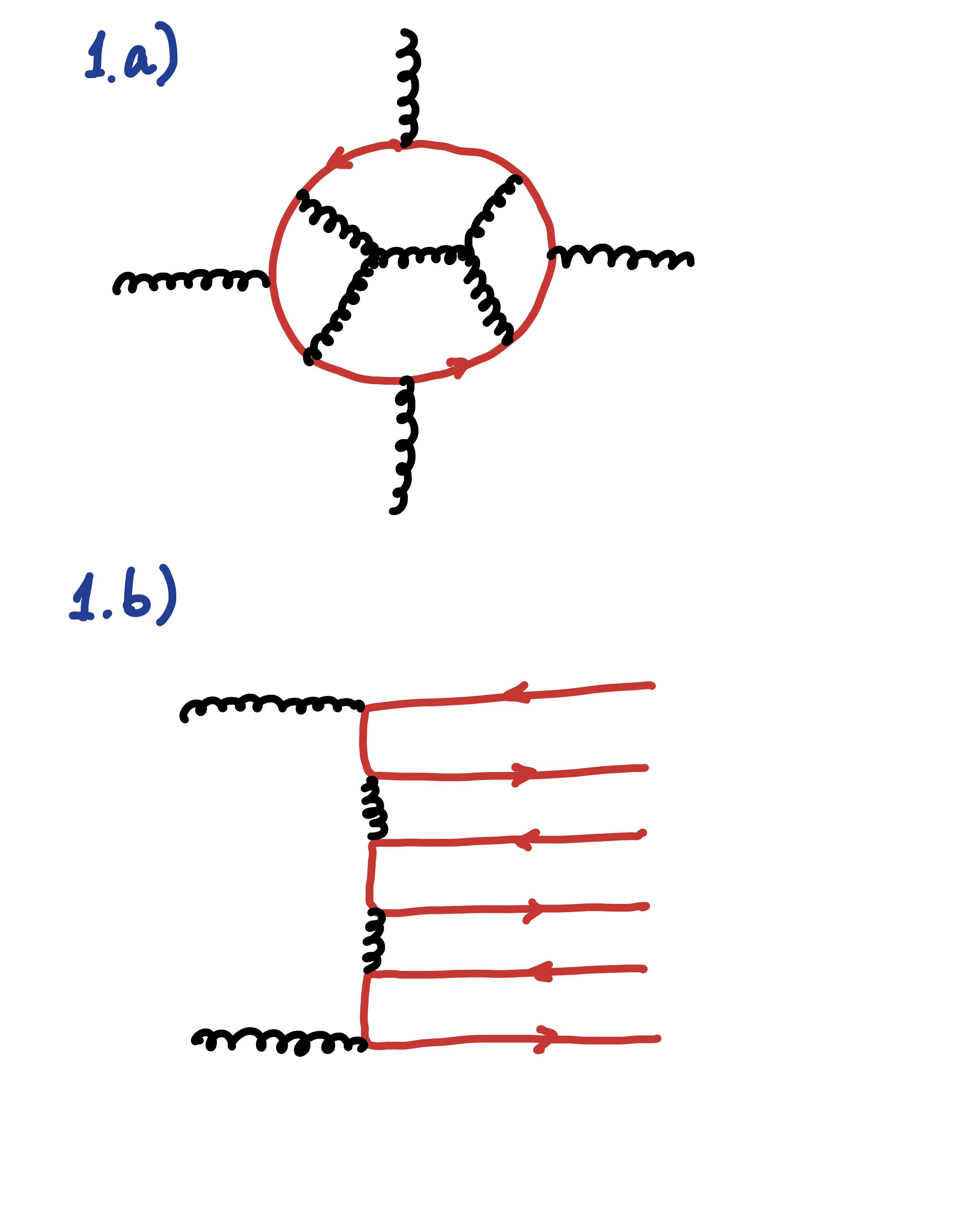}
 		\caption{ {\bf 1.a)} An example of a quantum gravitational 
		amplitude that vanishes in the double-scaling limit
		 (\ref{PLimit}).  The gravitons are denotes by black wavy lines 
		and the species by solid red ones.  This amplitude scales as $\alpha^5N_{\rm sp} = \lambda \alpha^4$ and is zero for (\ref{PLimit}). 
{\bf 1.b)} A process of graviton annihilation into $n$ species.   For $n \sim N_{\rm sp}$ the cross section is non-zero and for $\lambda \sim 1$ can even saturate unitarity due to a maximal entropy of the final state. It therefore provides an example of a quantum gravitational process that survives in the limit (\ref{PLimit}). } 	
\label{processes}
 	\end{center}
 \end{figure}

We now take 
the double-scaling limit  (\ref{SP1}). In this regime, both $N_{\rm sp}$ and $M_P$ become infinite while 
  the scale $M_*$ is kept finite.   We shall refer to this as the {\it species 
  limit}.  
  It is clear from (\ref{AlphaM}) that in this regime 
$\alpha$ vanishes while 
  $\lambda$ remains finite.   The behaviour of various parameters 
  in the species limit is summarized as, 
    \begin{eqnarray} \label{PLimit} 
&& M_P \rightarrow \infty\,,~ N_{\rm sp} \rightarrow \infty\,,~\alpha \rightarrow 0\,, \\ \nonumber 
&& M_* = {\rm finite}\,, ~ \lambda = {\rm finite}\,. 
\end{eqnarray}
In this regime, the quantum gravity simplifies substantially.  
All quantum 
gravitational processes, in which each power of $\alpha$ 
is not accompanied by $N_{\rm sp}$, vanish.
 An example is provided by Fig.\ref{processes}a. 
\\

Let us consider the theory of Einstein gravity,
minimally coupled to $N_{\rm sp}$ 
particle species.   We shall treat it as an effective  QFT  
of a graviton ($h_{\mu\nu}$) and of other particle species, formulated on an $S$-matrix vacuum 
of Minkowski.  In such a theory, the linearized metric perturbations
are described by excitations of a canonically normalized graviton field around the Minkowski space,  
$\delta g_{\mu\nu} = \eta_{\mu\nu} +  \frac{h_{\mu\nu}}{M_P}$. 
Likewise, the states of arbitrary complexity are viewed as excited states 
constructed on top of the Minkowski vacuum.  
 For definiteness, we assume that the species are light 
and stable. 
We now take the limit (\ref{PLimit}). \\

Let us first consider only the states of finite energies. 
That is, the total energy is allowed to be arbitrarily large, 
as long as it is finite. 
Obviously, in such states, the occupation numbers of finite 
energy (frequency) quanta are also finite.  \\

In such a case, the theory effectively 
reduces to a linear theory of a massless graviton $h_{\mu\nu}$,
coupled to species. The Lagrangian can be written as, 
 \begin{equation} \label{Einstein} 
 {\mathcal L} =   h^{\mu\nu} {\mathcal E}_{\mu\nu}\, 
 +  \, \frac{1}{M_*\sqrt{N_{\rm sp}}} h_{\mu\nu} \sum_{j=1}^{N_{\rm sp}}
  T_{j}^{\mu\nu}   
+  \,...\,,
\end{equation}
 where ${\mathcal E}_{\mu\nu}$ is the linearized Einstein tensor
 and $j=1,2,...,N_{\rm sp}$ is the species label.  
 $T_{j}^{\mu\nu}$ are their energy momentum tensors. 
 All the self-couplings of the graviton vanish. The same is true 
about its  
 couplings to species that are higher than quadratic in $h_{\mu\nu}$. \\

 At the loop level, a relevant correction is a renormalization of
 the graviton kinetic term by species. In the parameterization of 
 (\ref{Einstein}), this correction is of zeroth order 
 in $1/N_{\rm sp}$ and, therefore, survives in the species limit (\ref{PLimit}). \\
 
  On the other hand, the corrections to non-linear 
couplings of gravitons vanish. 
Correspondingly, the same applies to any transition amplitude between finite numbers of 
initial and final gravitons,
\begin{equation} \label{INF}
N_{in} \rightarrow  N_{f}\,. 
\end{equation} 
Such amplitudes are zero to all orders within the validity 
of $1/N_{\rm sp}$-expansion.  An example of a four-loop correction to
$2 \rightarrow  2$ graviton scattering amplitude, is provided by 
Fig.\ref{processes}a. This correction scales as $\alpha^5N_{\rm sp} = \lambda \alpha^4$ and is zero for (\ref{PLimit}). 
  \\
 
 Notice,  for the suppression of higher order corrections, the 
 relation (\ref{Mstar}) is of fundamental importance. It is the ability 
 of species, to lower the cutoff relative to $M_P$, that enables us to 
 consistently restrict the momentum-transfer among the particles 
 (both, virtual and real) by the scale $M_*$.  Correspondingly, this
 restricts the coupling $\alpha$ by $1/N_{\rm sp}$, as 
 indicated by (\ref{AlphaM}). \\

  We also wish to remark that starting with a minimal version of 
 Einstein-Hilbert action, is not essential. We could have 
 instead started with a theory in which infinite series of high order 
 curvature invariants are preemptively added to this action.
 An each {\it additional} derivative would be accompanied by the 
 factor $1/M_*$.   After canonically normalizing the fields, 
 an $n$-graviton interaction vertex is suppressed by 
$1/M_P^{n-2}$.  Due to this, the expansion in terms of the curvature invariants, would effectively be converted 
 into a $1/\sqrt{N_{\rm sp}}$-expansion in terms of graviton interactions. 
  Consequently, 
all nonlinear self-interactions of graviton would again vanish in the species limit (\ref{PLimit}).  \\

 The simplification of quantum gravity in 
 the species regime (\ref{PLimit}), allows us to extract 
 non-perturbative 
 processes.   
 As already pointed out in \cite{Dvali:2020wqi}, in this limit certain collective processes are magnified to the level of saturating unitarity.  
 One example considered there is the process,
\begin{equation} \label{2inN}
 {\rm graviton + graviton} ~\rightarrow ~ N-{\rm particles} \,,
 \end{equation}     
 in which two gravitons, of center of mass 
 energy $E = Nq$, annihilate in $N$ quanta of various species
 \footnote{Of course, since the final state quanta gravitate, they must 
 be properly dressed by soft gravitons. 
 This standard IR-dressing is independent of the entropy of species
 and is assumed to be done in all processes of interest.}.
 An each final particle,  carries the energy $q$, which is 
 a $1/N$-fraction of the initial energy. 
 A  particular diagram contributing to such a process is given by 
 Fig.\ref{processes}b.  The amplitude of this process is proportional to $\alpha^{N/2}$ where the effective coupling $\alpha$ 
 is given by  (\ref{Alpha}).  Thus, it is of order $N/2$ in $\alpha$-expansion. 
 \\
 
 The saturation regime is established when the occupation number 
 in the final state is $N=1/\alpha$ while $N_{\rm sp}$
 approaches the same value from below.
 Starting from $\lambda \ll 1$  ($N_{\rm sp} \ll N$), 
 we can scan the process by gradually increasing $\lambda$,
 keeping other parameters fixed, until it saturates
 unitarity. \\
    
 Usually, such a process would be expected to exhibit an exponential suppression, 
 ${\rm e}^{-\frac{1}{\alpha}}$,  characteristic of non-perturbative amplitudes. And indeed, it would, if the number of species were small. 
But, in the present case the story is different.  
For (\ref{PLimit}), the transition probability is enhanced by the 
 exponentially large number of possible final states, since 
 the quanta can come from various combinations of species.   \\
 
  This multiplicity can be quantified by the entropy 
 $S$, which can be expressed as a function 
 of $\alpha$ and $\lambda$.
 Its generic form is, 
 \begin{equation} \label{ent}
   S = \frac{1}{\alpha} \ln (f(\lambda)) \,,
 \end{equation}  
where $f(\lambda)$ is a function of  $\lambda$. 
It satisfies $f(\lambda) ={\rm e}$ for a certain critical value
$\lambda = \lambda_*$, typically order one. 
The precise form of the function $f(\lambda)$ is 
representation-dependent and is unimportant for the present discussion. 
For the intuitive usefulness, we just say that its typical form 
is   
 \begin{equation} \label{ent}
   f(\lambda) = (1 + 1/\lambda)^{c_1\lambda}
   (1 + \lambda)^{c_2}{\rm e}^{-c_3} \,,
 \end{equation}  
where $c_1,c_2,c_3$ are positive numbers of order one.    
   More explicit forms can be found
 in  \cite{Dvali:2020wqi}. \\

 Now, near the saturation point, the cross section 
 (measured in units of $1/q$)
 takes the form 
 \begin{equation} \label{crossA} 
 \sigma = {\rm e}^{-\frac{1}{\alpha} + S} \,.    
 \end{equation} 
 Notice that, thanks to the limit (\ref{PLimit}), the 
 saturation point is insensitive to a non-exponential 
 pre-factor which, therefore, 
 can be set equal to one.  The error committed in this way 
 is of order $\alpha$ and vanishes in the limit (\ref{PLimit}). \\ 
  
From (\ref{crossA}) it is clear that  the saturation takes place for 
 \begin{equation} \label{satsat} 
 S = \frac{1}{\alpha} \,.  
 \end{equation} 
As already said, this is reached for $\lambda =\lambda_* \sim 1$. 
It is clear that at this point, the final state 
of species satisfies a saturon relation (\ref{saturation}). 
Notice, for $q = M_*$, the unitarity of the process
bounds $N_{\rm sp}$ exactly by (\ref{Mstar}).  This gives an alternative derivation  
of the species bound on $M_*$ from saturation.   \\

\subsection{Collective Limit}

 We have seen that the species limit (\ref{PLimit})  
allows to cleanly extract a set of non-perturbative processes. 
This is possible, despite the fact that the gravitational coupling 
$\alpha$ vanishes. 
 In the considered process, the suppression is compensated 
 by the number of species and 
 the corresponding entropy of the state.  
 We shall now turn to a different class of quantum gravitational  processes that survive in the species limit (\ref{PLimit}). 
 These are the transitions that are  
enhanced by the occupation number of gravitons, $N$.  \\ 

In such cases,  an interaction vertex of gravitons that vanishes 
in the limit (\ref{PLimit}), can nevertheless  generate a  
non-trivial transition.  For example, a $2\rightarrow 2$ transition amplitude, 
caused by a tree-level four-graviton vertex, is suppressed by 
$\alpha$. Nevertheless, the transition matrix element  
can be non-zero in the limit (\ref{PLimit}), provided the occupation number of quanta scales as, 
\begin{equation} \label{N1/A} 
    N \propto \frac{1}{\alpha} \,.
\end{equation}
In order to account for such cases, 
it is convenient to 
introduce the notion of the {\it collective coupling}, defined as, 
\begin{equation} \label{Col}
     \lambda_c \equiv  N \alpha \,.
\end{equation} 
Originally, this concept was introduced in 
the context of the black hole $N$-portrait.\\ 

It is important not to confuse the occupation number 
of gravitons $N$ in a state, with the number of species $N_{\rm sp}$ in the theory.  Correspondingly, we should not confuse the collective coupling 
$\lambda_c$, with the species coupling 
$\lambda$ defined in (\ref{Lambda}).   
 The species coupling $\lambda$, is a parameter  
of the theory, since it depends on $N_{\rm sp}$.  
In contrast, the collective coupling $\lambda_c$, is a parameter  
of the state, which depends on the occupation number 
 $N$ in that state. \\ 

 The concept of collective coupling, allows us to define the second double-scaling limit, 
 \begin{eqnarray}
 \label{CLimit}  
 && M_P \rightarrow \infty\,, ~ N \rightarrow \infty\,,~\alpha \rightarrow 0\,, \\ \nonumber
  && \lambda_c = {\rm finite}\,, ~ N_{\rm sp} = {\rm finite} \,.
 \end{eqnarray}
We shall refer to it as the {\it collective limit}. 
This limit requires some clarification. Since we kept 
$N_{\rm sp}$ finite, the fixed hierarchy between the cutoff 
 $M_*$ and $M_P$ is maintained.  In such a case, setting $\alpha$ to zero implies a restriction on $q$, for example, 
 by considering a tree-level process with a finite 
 momentum transfer among gravitons.  
 Such restrictions need not be obeyed by the virtual quanta. 
 As a result, the corrections 
 from UV-sensitive loops, translate as $1/N_{\rm sp}$-corrections.
 This is different from the  species regime (\ref{PLimit}), in which the 
 analpogous corrections vanish. 
  \\

Obviously, in the regime (\ref{CLimit}),
the species coupling $\lambda$ vanishes together with 
$\alpha$.  Because of this, we may expect that  
quantum gravity becomes trivial in IR \cite{Brustein:2009ex}.
However, this expectation does not capture the collective effects 
that are enhanced by $N$.  
 \\

Naturally, those processes that are controlled by the collective coupling 
$\lambda_c$, are non-zero in the limit (\ref{CLimit}).  
  For example, we choose an initial state $\ket{N}$, in which
 gravitons of certain wavelength $R$ are macroscopically 
 occupied, so that $\lambda_c$ is non-zero. 
  Let us consider a process, in which two initial gravitons   
 re-scatter into a pair of final gravitons, as it is sketched in 
 Fig.\ref{NtoN1}b. Alternatively, the same pair can annihilate 
 into a pair of  species, as described by Fig.\ref{NtoN1}a. 
 The matrix elements of both transitions are proportional to 
$\lambda_c$. Correspondingly, they are both non-vanishing in the collective 
limit (\ref{CLimit}).
At the same time, all the higher order corrections in $\alpha$, 
not accompanied by the corresponding powers of $N$, vanish.  
 \\

The processes of the sort of Fig.\ref{NtoN1}, can have a wide range of physical applications.  In particular, they describe the processes of the gravitational particle-creation by black holes and by de Sitter in the corpuscular theory of \cite{Dvali:2011aa, Dvali:2013eja}. \\

The sketch in Fig.\ref{NtoN1}b, also reveals why the self-binding potential 
of $N$ overlapping gravitons,  
 survives in the collective limit (\ref{CLimit}). 
This diagram describes the origin of the attractive potential, 
at the level of a virtual one-graviton exchange among the
real gravitons.   
 The attractive potential energy, experienced by each graviton  
from its $N$ neighbours, is $E_{\rm pot} \sim \lambda_c/R$.  
For $\lambda_c \sim 1$,  this potential can balance the kinetic energy of a would-be free 
graviton ($E_{\rm kin} \sim 1/R$).   At this point, 
$N$ gravitons form a self-sustained bound-state. 
This bound-state represents a corpuscular description of a black hole 
\cite{Dvali:2011aa, Dvali:2012en, Dvali:2013eja}.
  
 \subsection{Species Magnifier Limit} 
 
 Finally, we shall discuss a regime,  which is of special interest 
 for our applications. This regime was previously introduced in 
 \cite{Dvali:2020wqi} and we shall refer to it as the {\it limit of species magnifier}.  It has the following form, 
   \begin{eqnarray} \label{MLimit}
&& \alpha \rightarrow 0\,,~N \rightarrow \infty\,,~ N_{\rm sp} 
 \rightarrow \infty\,, \\ \nonumber
 &&M_* = {\rm finite}\,, ~  \lambda = {\rm finite}\,, ~\lambda_c  = {\rm finite} \,.
  \end{eqnarray}
Basically, the above limit describes the species regime 
(\ref{PLimit}), in the presence of a state with 
infinite occupation number $N$ of gravitons, such that   
the ratio $N/N_{\rm sp}$ is finite.   
 \\
 
 Correspondingly, while this regime maintains benefits 
 of the species limit (\ref{PLimit}), it enables to
 expose non-trivial collective phenomena.  For example, 
 a four-graviton interaction vertex can trigger 
 a non-trivial transition from an initial $N$-graviton state via a re-scattering of two constituents. 
  The matrix 
 element of this process is controlled by $\lambda_c$. 
 At the same time, the loop corrections to the same vertex, such as 
 Fig.\ref{processes}a, vanish. \\

In order to capture more essence of the limit (\ref{MLimit}), let us compare 
the outcomes of the process of Fig.\ref{NtoN1}a in the regimes
(\ref{CLimit}) and (\ref{MLimit}).    
  As already mentioned, this process describes the 
  scattering of two gravitons into an arbitrary pair of species. 
  The rate of this process is enhanced by $N_{\rm sp}$ due to the number 
 of available final states. It scales as, 
 \begin{equation} \label{2to2S}
   \Gamma \sim q \lambda_c^2 N_{\rm sp} \,,
 \end{equation} 
 where $q = 1/R$ is a characteristic energy of the initial gravitons.  
 The inverse of this rate, 
  gives the time-scale
   during which the number of initial 
  gravitons is reduced roughly by two. Of course, there are other processes that contribute but this one suffices  
  for making the point.  \\
  
  As a reference unit of time, we can adopt $R = 1/q$.
 During this time,  the relative reduction of the graviton number is, 
  \begin{equation} \label{BR}
   \frac{\Delta N}{N}  \sim 
    \frac{N_{\rm sp}}{N} \,  \lambda_c^2\,.
 \end{equation}  
This quantity represents a measure of the quantum gravitational back reaction on the state. \\
  
  Correspondingly,  the time-scale $t_{\rm half}$, 
   during which the initial state would loose roughly half of  
 its constituents, is given by, 
  \begin{equation} \label{life2}
   t_{\rm half}  \sim \frac{R}{\lambda_c^2} \frac{N}{N_{\rm sp}} \,.
 \end{equation} 
 We shall now compare the behaviour of the above three quantities 
 (\ref{2to2S})-(\ref{BR})-(\ref{life2}), in the two different regimes of 
 (\ref{CLimit}) and (\ref{MLimit}). \\
 
 We start with the collective limit (\ref{CLimit}). In this regime, 
 the emission rate (\ref{2to2S}) is finite. However, the quantum gravitational back reaction (\ref{BR}) is zero. Correspondingly,
 the half-decay time (\ref{MLimit}) is infinite. The physical reason behind this 
 behaviour is no secret.  In the collective limit (\ref{CLimit}), the 
 initial graviton occupation number $N$, is infinite. Such a state behaves as a graviton reservoir of infinite capacity. 
 A finite depletion rate, is not capable of 
 destroying such a system within any finite time.  In other words,
 the fraction of gravitons depleted within any finite interval of time, is of measure-zero.  This explains why the back reaction (\ref{BR}) is vanishing. \\
 
 In the species magnifier regime (\ref{MLimit}), the situation changes drastically. 
 The infinite number of species $N_{\rm sp}$, opens up an infinite 
 number of the decay channels.  Correspondingly, the rate (\ref{2to2S}) blows up. However, the problem is regular. The back-reaction 
 (\ref{BR}) is finite and so is the half-decay time 
 (\ref{life2}) of the graviton 
 condensate.  \\
 
 The important aspect of the magnifier limit is that it 
 captures the quantum gravitational 
 back-reaction, despite the fact that $\alpha=0$.  
 At the same time,
 the secondary effects, that are higher order in $\alpha$, 
 vanish. \\

  To summarize,  working in the regime (\ref{MLimit}) provides us with a double advantage.   
  First, it allows to cleanly distill some collective and non-perturbative quantum gravitational phenomena.  Secondly, it shows that, for large 
$N_{\rm sp}$, the quantum gravity can leave significant imprints 
at distances much larger than the Planck length.   
 The species literally act as a ``magnifying glass" for quantum gravity. \\

  In what follows, we shall apply the species regime 
for extracting certain quantum gravitational effects in de Sitter and in black holes and for understanding their observational implications.

 \section{Corpuscular Theory}  \label{CP}
 
 \subsection{Motivation} 
 
 As we have seen, 
 the magnifying effect of species can be deduced already within the semi-classical treatment.  The relations such as (\ref{Mstar}),  
 do not require the knowledge of an internal corpuscular structure of either a black hole 
 or a de Sitter. 
  The semi-classical picture is of course incomplete.
  Although this picture allows to derive the species bound (\ref{Mstar}),   
 it cannot grasp the fundamental physics behind it.  
 For example, it is not clear, how a back-reaction from 
 particle-creation, 
 affects 
 a black hole or a de Sitter.  
  \\
 
 In order to answer the next layer of questions, we  
 must work within a more fundamental framework, in which the corpuscular structure 
 is visible.  Once we have such a picture, we shall 
 supplement it with a large number of light particle species. 
 This shall enable us to work in the regime (\ref{MLimit}) and 
 to use species 
 as a magnifying tool.  \\

 The corpuscular picture we shall work with, is the quantum $N$-portrait 
  \cite{Dvali:2011aa, Dvali:2012en, Dvali:2013eja}. 
In this picture,  a black hole  
 or a de Sitter Hubble patch, is described as a composite state of 
 {\it soft}  gravitons of mean occupation number $N$. 
 This state satisfies the saturation relation (\ref{saturation}). \\
 
 A particularly strong motivation
 for the corpuscular picture, is provided  
 by the $S$-matrix formulation of quantum gravity.
 This formulation demands the existence of a 
 valid $S$-matrix vacuum. All other states, must be described as
 excited states, constructed on top of this vacuum.  \\

 The de Sitter cannot serve as a valid $S$-matrix vacuum for 
 quantum gravity. The absence of a globally-defined time, 
 is part of the problem. However, the issue is more subtle and is linked with the very nature of quantum gravity/string theory \cite{Dvali:2020etd}.
 The point is that, in general, a valid vacuum must not recoil and absorb information 
in a scattering process. 
A de Sitter can satisfy this requirement, but exclusively in the following 
double scaling limit,
\begin{equation} \label{doublescale}  
 M_P \rightarrow \infty,~R={\rm finite} \,. 
\end{equation}
In this limit, the quantum graviton coupling  (\ref{Alpha}), vanishes. 
Hence, whenever de Sitter becomes a rigid vacuum,  gravitons (closed strings) 
decouple. In other words,  de Sitter can be promoted into a ``vacuum" 
only at the expense  of trivializing the 
 graviton (closed string)  $S$-matrix. 
 Notice,  at the same time, non-gravitational interactions can remain 
 non-trivial for (\ref{doublescale}).  In particular, for interactions that are asymptotically-free,  
 an approximate $S$-matrix treatment becomes better and better 
 at short distances and can be consistently formulated in the 
 limit (\ref{doublescale}). 
 \footnote{In this sense, there is no {\it a priory}  inconsistency 
 in an approximate $S$-matrix treatment on a temporary
 cosmological backgrounds (for various examples, see, e.g., \cite{Gorbenko:2019rza},  \cite{Pajer:2020wnj}). }
  \\

 The important message we take from this analysis, is that 
 the $S$-matrix formulation excludes  a de Sitter vacuum for any finite value of $M_P$.  
 The only remaining option is to treat de Sitter 
 as an excited (coherent) state constructed on top of a valid $S$-matrix vacuum such as Minkowski \cite{Dvali:2011aa, Dvali:2013eja, Dvali:2014gua, Dvali:2017eba}.  
 Various other aspects of this idea have been discussed 
 \cite{Kuhnel:2014gja, Kuhnel:2015yka, Casadio:2015xva, Berezhiani:2016grw}, including string context \cite{Brahma:2020htg}.
  \\
 
  The $S$-matrix justification for the corpuscular structure of a 
  black hole \cite{Dvali:2011aa}, is more straightforward.  
  There is a little doubt that a black hole 
  of size $R$,   
  can be produced in an $S$-matrix process of center 
 of mass energy $E \sim M_P^2R$ \cite{tHooft:1987vrq, Amati:1987wq, Gross:1987kza}.  
 The new insight, brought by the corpuscular $N$-portrait \cite{Dvali:2011aa} 
is the identification of the relevant $S$-matrix process 
in form of $2\rightarrow N$ graviton scattering,  
where the occupation number of the final gravitons is  $N \sim (RM_P)^2$ and their wavelengths are $\sim R$.  
It has been observed \cite{Dvali:2014ila, Addazi:2016ksu} 
that such a $2\rightarrow N$ graviton scattering process, saturates unitarity precisely when the entropy of the final 
state matches the  Bekenstein-Hawking entropy of a black hole 
of size $R$.   
  It is therefore natural to describe a black hole
 as a ``loose" bound-state 
  of $N$ such quanta.  This justifies the idea of the black hole 
  $N$-portrait, from the $S$-matrix perspective.  \\

\subsection{Two Types of Constituents}

 The corpuscular pictures
for a black hole and for de Sitter exhibit close similarities. 
Therefore, we shall discuss them in a common language.  
This language also allows to grasp the fundamental differences 
between the two systems. These differences, become important 
at the later stages of their time-evolutions. \\

In both cases, the relevant gravitational radius shall be denoted by $R$.  In classical theory, 
this scale describes a Hubble radius  
($R=1/H$) for de Sitter and a Schwarzschild radius for a black hole. \\

The constituents are divided in two main categories: 
The {\it master modes} and the {\it memory modes}. 
\\

 The {\it master modes}, to be symbolically denoted by 
 creation/annihilation operators $\hat{a}^{\dagger}$/$\hat{a}$,
 are the main contributors 
 into the energy of the system (a black hole or a Hubble patch).
 The diversity of the master modes is small, but they come in high 
 occupation numbers.  Because of this, they contribute predominantly 
 into the 
 energy but very little into the information capacity of the system. 
 \\
 
 The {\it memory modes}, to be denoted by 
 $\hat{b}^{\dagger}$/$\hat{b}$, play the opposite role. 
 They are responsible for the information storage capacity of the system.
 This capacity is measured by the micro-state entropy $S$. 
 This stands for Bekenstein-Hawking entropy for a black hole and Gibbons-Hawking entropy for de Sitter.  The memory modes are initially (nearly) gapless.  Due to this, they contribute negligibly into the energy. 
 However, unlike the master modes, the memory modes come 
 in a very large diversity of ``flavors". This is the reason for why they account for the entropy of the system.  
  \\
 
   We shall briefly review each category.
 
  \subsubsection{Master Modes}

  The characteristic frequencies 
  (${\mathcal E}_{\rm master}$) and wavelengths
  ($l_{master}$) of the master   
modes are both set by the scale $R$,  
\begin{equation} \label{GMaster}
{\mathcal E}_{\rm master}  \sim \frac{1}{l_{\rm master}} \sim R  \,. 
\end{equation}
That is, the dispersion relations  of the master modes are not 
too far from the dispersion relations of the free gravitons, propagating 
on top of the Minkowski vacuum. \\

This is explained by the self-sustainability condition that was already discussed 
previously.  The attractive potential, experienced by each master 
graviton,  
is comparable to its kinetic energy.  
Therefore, the collective gravitational interaction, while sufficiently strong for holding the master modes together, only mildly modifies their dispersion relations. 
 The essence of this collective interaction was already discussed and illustrated by a sketch in Fig.\ref{NtoN1}b. \\
 
 The master gravitons are the main contributors 
 into the initial energy of the system.  An each quantum contributes  
 roughly $1/R$.  The total energy is thus proportional to their 
 occupation number, 
 \begin{equation}\label{EM} 
 E= N{\mathcal E}_{\rm master} = \frac{N}{R} \,.
 \end{equation} 
 Given the standard relation,
 \begin{equation}\label{BHM} 
 E= RM_P^2 \,, 
 \end{equation} 
  between the energy (mass)  and a size 
 of a black hole (or a Hubble patch), we get, 
  \begin{equation}\label{NNN} 
N = (M_PR)^2 \,.
 \end{equation}   
  The basic quantum gravitational coupling 
of the master gravitons is given by (\ref{Alpha}) evaluated 
for $q \sim 1/R$, 
 \begin{equation} \label{AlphaDS} 
 \alpha = \frac{1}{(RM_P)^2}\,.    
 \end{equation} 
 This given two important relations. 
 First, for both systems the above expression is equal to the 
inverse of the entropy $S$ (Bekenstein-Hawking entropy for a black hole and Gibbons-Hawking entropy for de Sitter). 
 Both entropies are equal to  (\ref{BH}). \\

 Secondly, the number of constituents  $N$ is 
 equal 
 to the inverse of their coupling (and equivalently, to the entropy). 
 Thus, both, a black hole and a de Sitter, satisfy the saturation 
  relation (\ref{saturation}). Taking into account 
  the explicit form of the entropy (\ref{BH}), the saturation relation reads as, 
 \begin{equation} \label{sat} 
 N  = \frac{1}{\alpha} = S = (RM_P)^2\,.    
 \end{equation}
 As pointed out in \cite{Dvali:2020wqi, Dvali:2019jjw}
 and also explained earlier in the present paper, the area form of the entropy 
 (\ref{BH}) represents a particular case of the generic property of saturons. 
The entropy of a saturon is always given by its area measured 
in units of the decay constant of the Goldstone boson of spontaneously broken Poincare symmetry. 
 The emergence of such a Goldstone boson is generic, because any saturon breaks Poincare symmetry spontaneously. In case of a black hole or a de Sitter, the Goldstone boson in question is the graviton, but the effect is universal and goes beyond gravity (See, \cite{Dvali:2020wqi, Dvali:2019jjw} for explicit examples.) \\
 
  The concept of spontaneous breaking of Poincare symmetry 
  by the  saturated $N$-graviton state, 
is best defined in large-$N$ limit (\ref{CLimit}). 
The order parameter of this breaking is
 $\sqrt{N}/R$, which,  by Goldstone theorem, determines the decay constant of a canonically normalized Goldstone mode.  
 As it is clear from (\ref{NNN}), this quantity is nothing  but  the Planck mass, $M_P$, which represents the graviton decay
constant.  This fully matches the fact that the Goldstone boson
of Poincare symmetry, originates from the graviton.  
The above  explains why $M_P$ enters in the expression
(\ref{BH}) for the area-form of the entropy.   
  \\
 
 The relation (\ref{sat}) indicates that both
 systems  (black hole and de Sitter) represent saturons.  
This is defining for their physical properties. 
 
 \subsubsection{Memory Modes}

 Unlike the master modes, the memory modes 
have very short wavelengths (of order the cutoff) 
 but almost zero frequencies. Namely, the
 energy gap for exciting a memory mode is, 
 \begin{equation} \label{Gap}
  {\mathcal E}_{\rm memory}  \sim \frac{1}{NR}  \ll \frac{1}{l_{\rm memory}}\,. 
 \end{equation} 
  Obviously, the dispersion relations of the memory modes 
 are very different from those of free gravitons.  A free graviton, of 
 some wavelength $l$, propagating on   
 Minkowski vacuum,  would 
 have a frequency, $\sim 1/l$.  This is due to the 
 Poincare symmetry of the Minkowski background. \\
 
 However, the dispersion relations of the memory modes, are affected 
 by the presence of the master modes.  
 As already discussed, the state of $N$ master modes, breaks the Poincare symmetry spontaneously. 
 As a result of this breaking, the dispersion relations of the would-be 
 ``hard" memory modes are strongly affected. Their wavelengths stay short, 
 but their frequencies are redshifted to almost zero.
 In fact, the memory modes become exactly gapless in 
 the collective limit (\ref{CLimit}).  At finite $N$, their gaps 
 scale as  (\ref{Gap}). \\
 
 The variety of the ``flavors" of the memory modes is of order 
 $1/\alpha$. This is determined by the number 
 of different momentum oscillators that become gapless at the 
 saturation point.    
 Due to this,  the memory modes are the main contributors into the entropy of the system. 
 These modes are gapless because of the critical occupation 
number of the master modes. That is, the energy gaps of the
memory modes depend on the occupation number of the master mode.
The gaps collapse to (almost) zero  when the occupation 
number of the master mode reaches a critical value.\\

 This phenomenon, 
 schematically, 
can be described by the following Hamiltonian 
(see, e.g., \cite{Dvali:2017nis, Dvali:2018xpy}), 
\begin{equation} \label{Hgap}
 \hat{H} =  \sum_b  \omega_b \left (1-\alpha \, \hat{a}^{\dagger}\hat{a}\right )^2 
 \hat{b}^{\dagger}\hat{b} \,.
\end{equation} 
For simplicity of illustration, we took a single master mode. 
The summation is over various memory modes. 
$\omega_b$ are their would-be frequencies in the
Minkowski vacuum, 
which typically are very high (of order the cutoff or so). \\

The reader should not be alarmed by the oversimplified
appearance of the 
above Hamiltonian. In particular, by its number-conserving form. 
These details are unimportant and have been neglected for 
making the essence of the phenomenon maximally transparent. 
Thanks to the power of saturation, the 
(\ref{Hgap}) captures this essence surprisingly well.  
 In fact, near the saturation point (\ref{sat}),
 the dynamics of a generic system effectively reduced to the above,  
 up to a proper Bogoliubov transformation.   
\\

In a state with the occupation number of the master mode 
$\langle \hat{a}^{\dagger}\hat{a} \rangle \equiv  N$, the  
effective energy gaps of the memory modes are given by 
\begin{equation} \label{Hgap1}
 {\mathcal E}_{\rm memory} =  \omega_b \left (1-\alpha N \right )^2  \,. 
 \end{equation} 
It is clear that the memory modes become 
effectively gapless over a state 
in which $N$ is critical,
\begin{equation} \label{CNA}
 N  = \frac{1}{\alpha} \,.
\end{equation} 
Since the memory modes are gapless, an exponentially 
large number of degenerate micro-states emerges. These states differ by 
the occupation numbers of the memory modes. 
This is the reason behind the maximal entropy of the saturated system. 
 \\

Thus, we can say that in a saturated state the master mode 
{\it assists} \cite{Dvali:2018tqi}
 the memory modes 
in becoming gapless. Away from saturation (\ref{sat}), the memory modes are very costly in energy, but at the  
saturation point,  they cost nothing.  Thanks to this phenomenon, the memory modes 
can store a large amount of information at almost zero energy 
expense. 
 \\

Thus, a black hole or a de Sitter state is described by the 
occupation number of  the master mode and the ones of the memory modes.  
 This state can be denoted as,
\begin{equation} \label{vec} 
  \ket{N} = \ket{N}_a \ket{n_1,n_2, ...}_b \,,  
\end{equation} 
where $\ket{N}_a$ is a state of a master mode, 
with occupation number $N$, and $\ket{n_1,n_2, ...}_b$ is a state of memory modes,
with various occupation numbers. 
 An each sequence, represents a {\it memory pattern}.  
 The number of distinct patterns 
 is exponentially large, 
 \begin{equation} \label{nst}
 n_{st} \sim {\rm e}^{1/\alpha} \,. 
 \end{equation}   
 As long as the system is saturated (\ref{sat}), the patterns are degenerate in energy (up to $1/N$ corrections).
They form a set of micro-states that  are classically indistinguishable. 
The corresponding micro-state entropy is, 
  \begin{equation} \label{SM} 
  S= \ln (n_{st}) = \frac{1}{\alpha} = (RM_P)^2 \,.
 \end{equation}
 This is the origin of  
Bekenstein-Hawking and Gibbons-Hawking entropies in the 
corpuscular picture. \\

As said, the above presentation captures the essence of the story. 
 We needed to display the necessary  
ingredients of the phenomenon before highlighting the magnifying role 
of the particle species in it.

\subsection{Saturation} 

The saturation relation (\ref{saturation}) can be written as,  
\begin{equation} \label{satA} 
{\rm Number}  = \frac{1}{\rm Coupling} = {\rm Entropy} \,.  
\end{equation} 
The saturons (the systems satisfying this relation) 
exhibit certain universal 
phenomena. For the present discussion, the most important is  
the effect of {\it quantum breaking} and 
the internal mechanisms leading to it. \\ 

 The main engine of the time-evolution of a saturated system, is 
 the re-scattering of the master modes. This re-scattering leads to 
 the processes that are observable on the time-scales $\sim R$. 
 For example, such is a process of particle-emission,  which leads to the decay 
 of a saturon.  \\

 Since the memory modes have very low frequencies 
 (\ref{Gap}),  the time-scale of their evolution would be, 
  \begin{equation} \label{tMEM} 
 t_{\rm memory}  \sim  N R \,.  
\end{equation} 
In addition, the memory modes are affected by the re-scatterings of the 
master modes, over a similar time-scale.  \\
 
There exist three important effects contributing into the time-evolution of the saturon state:

  \begin{itemize}

\item  The reduction of the occupation number of the master mode. This 
is the flip side of the particle emission. 

  \item {\it The inner (or self) entanglement}  \cite{Dvali:2013eja, Dvali:2014gua, Dvali:2017eba}:  
     The constituents of a black hole (or a de Sitter) become internally entangled among each other. 
     
  \item {\it The memory burden effect} 
  \cite{Dvali:2018xpy, Dvali:2018ytn, Dvali:2020wft}: 
  With the decrease of the occupation number of the 
  master mode, the system departs from saturation. 
 Correspondingly, the gaps of the memory modes grow. 
  This generates a quantum back-reaction force.

\end{itemize}
 
 All three effects lead to a complete breakdown of the semi-classical 
 picture. This phenomenon is referred to as the 
 {\it quantum breaking}. 
 The corresponding time-scale is called a 	{\it quantum break-time} 
$t_Q$. For a de Sitter and a black hole,  $t_Q$ was derived 
in  \cite{Dvali:2013eja, Dvali:2014gua, Dvali:2017eba}. For a generic saturated system
(without classical instability) 
 \cite{Dvali:2017eba, Dvali:2020wqi} it is given by 
(\ref{QQQ}), which can be presented as,  
   \begin{equation} \label{TQG} 
 t_Q  \sim  \frac{t_{\rm cl}}{\lambda} \,.  
\end{equation} 
Here, $t_{\rm cl}$ is the classical time-scale set by the frequencies
of the constituents, $t_{\rm cl} \sim R$. 
$\lambda \equiv  \alpha N_{\rm sp}$ is 
the analog of species coupling (\ref{Lambda}) ('t Hooft coupling), for the generic saturated system. \\   

 The various aspects of the above effects have already been studied  
 in saturated gravitational systems, such as black holes 
 and de Sitter. 
 In the present paper we are interested in the 
 magnifying role of species in these processes. For completeness, in parallel, we shall briefly reproduce the essence of the phenomena.

\subsection{Quantum Clock: Re-scattering and Depletion}

 Initially, the main engine of the time-evolution 
 is the re-scattering of the master modes. 
 The memory modes, since at this stage they are essentially gapless, 
 back react very little and can be ignored.  However, back reaction 
 becomes very important at later times. \\
 
\begin{figure}
 	\begin{center}
        \includegraphics[width=0.53\textwidth]{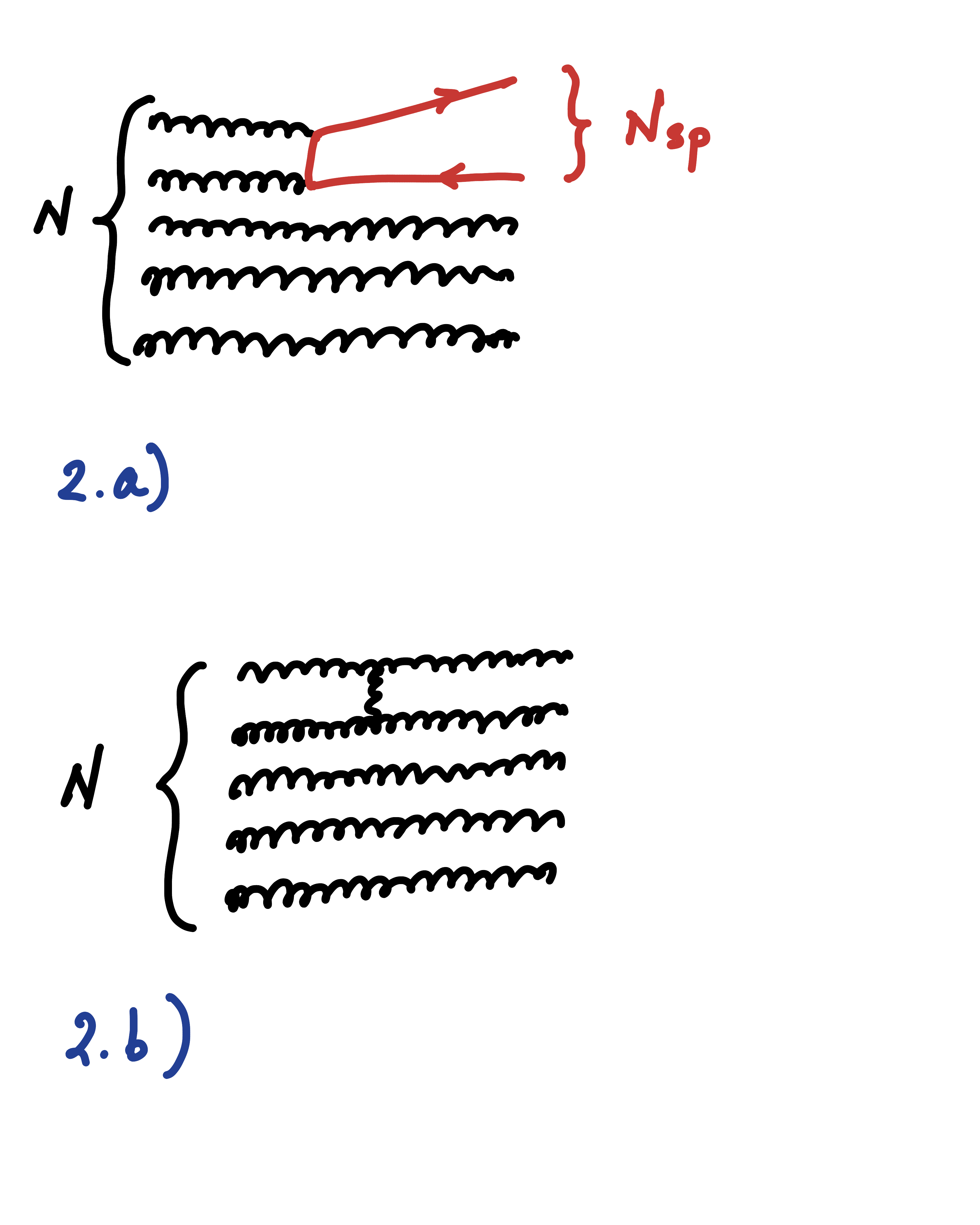}
 		\caption{ 2.a) Decay of the multi-graviton state into species (red line)
	illustrates the enhancement of the rate of the process
	due to high occupation number of the initial state.
	The suppression factor $\alpha^2$ is compensated by 
multiplicity  factor $N^2$.
 2.b) Illustrates the finiteness of the collective effect, 
 experienced by each graviton due to its interactions with the neighbours. 
 } 	
\label{NtoN1}
 	\end{center}
 \end{figure}

The initial re-scattering of the master modes, takes place regardless of the  
number of species in the theory.  However, species speed it up. 
 One outcome of this re-scattering, is the emission of free particles.
 The typical process is depicted in Fig.\ref{NtoN1}a. 
 For a black hole and a de Sitter, the produced particles
 describe Hawking and  Gibbons-Hawking quanta, respectively. \\
 
 For describing the process, we can use the formulas  
 (\ref{2to2S}), (\ref{BR}), (\ref{life2}) obtained 
 for a generic  $N$-graviton state.  The new specifics is that, 
 at initial times, the  parameters 
 are related by the saturation relation (\ref{sat}). \\ 

 The rate of particle-creation is, 
  \begin{equation} \label{rate} 
 \Gamma  = \frac{1}{R} (N \alpha)^2 N_{\rm sp} = 
 \frac{1}{R} N \lambda_c \lambda \,.    
 \end{equation} 
As expected, at initial times, 
the rate of particle-production, reproduces the semi-classical 
Hawking and Gibbons-Hawking rates. 
Note \cite{Dvali:2013eja}, a particular channel, describing a creation of a graviton,
corresponds to the standard tensor perturbation mode of inflation  
\cite{Starobinsky:1979ty}. \\

However, the corpuscular picture allows us to go beyond 
the semi-classical approximation.  Using the 
microscopic description, it is instructive to consider the physical meaning of particle-creation from different perspectives. \\

 First, using the saturation 
relation (\ref{sat}), valid for initial times, we can write the rate
(\ref{rate}) in terms of
the species coupling, 
 \begin{equation} \label{rateIN} 
 \Gamma_{in}  = \frac{1}{R} N_{\rm sp} = \frac{1}{R} \frac{\lambda}{\alpha} \,.    
 \end{equation} 
This form, clearly shows the magnifying power of species. 
As already discussed earlier for a generic $N$-state,  
in the double-scaling limit (\ref{MLimit}), the rate 
(\ref{rate})  
blows up.  However, this blow-up does not lead to any sort of inconsistency. 
 The saturation relation (\ref{sat}) guarantees that, in the same limit, 
the mass of a black hole (or of a de Sitter Hubble patch), 
(\ref{BHM}), becomes infinite. \\

 As a result,  the back-reaction over time $R$ (\ref{BR}), 
 is finite.  Due to saturation (\ref{sat}), in the present case, 
the back-reaction (\ref{BR})  takes the form,  
  \begin{equation} \label{BR1}
  \frac{\Delta N}{N} = \frac{N_{\rm sp}}{N} \,.   
\end{equation}
 Correspondingly,  the time of half-decay, 
  \begin{equation} \label{Half} 
t_{\rm half}  =  \frac{R}{\lambda} = \frac{R^3M_P^2}{N_{\rm sp}}
= R^3M_*^2 \,,     
 \end{equation} 
 is also finite. 
\\

The corpuscular picture makes it very clear that the particles are not produced ``for free". 
Due to back-reaction, the coherent state depletes and looses 
coherence.  This is true for black holes, as well as, for de Sitter. \\

The rate of the departure from coherence can be estimated by
using the fact that the number of quanta in a coherent superposition 
diminishes with roughly the same rate (\ref{rate}) as the particles are 
emitted. 
That is, 
 \begin{equation} \label{rateN} 
\dot{N} =  - \Gamma \,.    
 \end{equation} 
For estimating the back-reaction, 
in the leading order,  the coupling
$\alpha$ can be taken constant. Then, at initial times, the rate of 
particle-production changes as (as everywhere,
unimportant numerical coefficients are set equal to one),  
\begin{eqnarray}  \label{rateC} 
 \frac{\dot{\Gamma}}{\Gamma} &=& - \frac{1}{R} N \alpha^2  N_{\rm sp} = - \frac{1}{R} \lambda_c \lambda \,. 
 \end{eqnarray}
 Now, approximating the quantities on the right hand side by their 
 initial saturated values (\ref{sat}), we get, 
 \begin{equation} \label{rateS} 
 \Gamma(t)  = \Gamma_{in} {\rm e}^{- \frac{t}{t_{\rm half} }}  \,,
 \end{equation}   
 where $t_{\rm half}$ is given by (\ref{Half}). \\
 
 The expressions (\ref{rate}) and (\ref{rateS}) are instructive due to several reasons.
 First, for de Sitter, they reinforce the upper bound (\ref{BoundH}) on $H$. 
 We see from (\ref{rate}) that the violation of this bound would lead 
 to an unacceptably high rate of depletion. With such a gigantic 
 decay rate, the de Sitter state 
 would cease to exist in less than one Hubble time. 
 In full agreement with this, (\ref{rateS}) shows that the production rate 
 would drop in less than one Hubble time.  \\
 
 Similarly \cite{Dvali:2012uq},  for a black hole this reinforces the upper bound (\ref{Tstar}) on  Hawking temperature. A hotter black hole cannot exist. 
Such a black hole,  would emit all its 
constituents within the time less than its size. This would make no sense.
  \\

 To summarize,  
species produce the two inseparable effects. First, they enhance the rate of particle-production.  
 Secondly, as a result of back-reaction, species speed up an intrinsic 
 quantum clock.  This clock measures the depletion of 
 $N$-graviton state which  describes  
 a black hole or a de Sitter. The flip side of this process is the generation and growth 
 of entanglement which we shall now discuss.

\subsection{Entanglement and Page's Curve} 

The corpuscular picture reveals a new phenomenon, 
which we can call  
a {\it self-entanglement}  or an {\it inner entanglement} \cite{Dvali:2013eja}.  
 This takes place, both for de Sitter and for a black hole.  
In both cases, the growth of inner entanglement 
happens at the same rate as the particle emission. 
 Together with the depletion, 
the self-entanglement contributes into the departure from classicality. 
It reaches the maximum at quantum break-time 
\cite{Dvali:2013eja, Dvali:2014gua, Dvali:2017eba}, 
\begin{equation}\label{QBT}
  t_Q = \frac{R^3M_P^2}{N_{\rm sp}} = \frac{R}{\lambda} = R^3M_*^2 \,.
 \end{equation}
 It is easy to notice that this is a particular case 
 of (\ref{TQG}), applied to a gravitational saturon.
 The time-scale  (\ref{QBT})
 signals a complete break-down of the semi-classical picture. 
 The quantum break-time has been verified in various contexts
 \cite{Dvali:2017ruz, Kovtun:2020ndc, Kovtun:2020kcl, Berezhiani:2020pbv, Blumenhagen:2020doa}.  \\
 
 From (\ref{Half}) it is clear that  $t_Q$
matches the time of half-decay, but brings a new physical meaning. 
The effect is intrinsically 
corpuscular in nature and cannot be observed in semi-classical theory. 
This is why, unless additional assumptions are invoked, 
the behaviour of entanglement in semi-classical treatment 
is related with certain puzzling features of Page curve. 
According to Page \cite{Page:1993wv}, the entanglement must reach a maximal value after $t_{\rm half}$.   This is exactly what corpuscular theory predicts. 
The semi-classical picture cannot explain this. \\

   In semi-classical treatment, the sole engine for the Page's  
curve, is the entanglement between the 
black hole and an outgoing Hawking radiation. 
This creates a false impression that the entanglement grows 
indefinitely throughout the black hole existence. 
 However, this picture is incomplete and requires a corpuscular resolution which incorporates  the  inner entanglement
and memory burden effects. \\

  As already shown in \cite{Dvali:2013eja},   
 the corpuscular theory 
 immediately accommodates Page's central point. The 
 entanglement of a saturated $N$-particle state cannot grow indefinitely. 
 It reaches the maximum after the quantum break-time $t_Q$. 
 The fact that $t_Q$ 
 comes out equal to Page's time, can be taken as consistency check of 
 the picture.  \\

 The corpuscular approach shows that the analog of Page's time 
 exists also in de Sitter \cite{Dvali:2013eja}.  
 For both systems,  
 the re-scattering  of the constituent gravitons
 (which also leads to particle emission),
 generates an internal entanglement among the remaining constituents
 \cite{Dvali:2013eja}.
 Simultaneously, it creates the memory burden effect \cite{Dvali:2018xpy}
 in de Sitter \cite{Dvali:2018ytn} and 
 in black holes \cite{Dvali:2020wft}.

 \subsection{Inner entanglement} 
 
 After the emission of a single particle, the state 
 (\ref{vec})
 evolves into a following superposition, 
 \begin{equation} \label{Ent}
   \ket{N} \rightarrow \sum_j^{N_{\rm sp}} c_j\ket{N',j}\times\ket{j} \,,    
 \end{equation} 
 where $c_j$ are constants. 
Here, $j =1,2,...,N_{\rm sp}$ is the species label that characterizes various states 
of emitted quanta, which are denoted by $\ket{j}$. 
Since we are not in a semi-classical picture, the 
term  ``emitted quanta" requires a specification.
 It describes quanta with dispersion relations 
 that are close to quantum particles propagating in a background de Sitter (or Schwarzschild) metric.  \\
 
  Notice, in de Sitter, this is a fine definition at initial times. Later, towards the quantum break time, the classical de Sitter geometry looses the meaning.  By then,  the distinction between the emitted and retained quanta becomes 
 blurry. However, this is not an issue, since after $t_Q$ we are already in 
a quantum-broken phase. Beyond this point, the de Sitter cannot 
exist as a valid state. A graceful exit must take place 
before the time $t_Q$ elapses. \\

 In contrast, the black holes can exist beyond $t_Q$.  Also, for them, the notion of the emitted quantum is well defined at all times.  
\\

 Of course, at the initial times, 
the emission is democratic in species, since they are 
produced via graviton processes.  
  The states  
 $\ket{N',j}$ denote what remains from a black hole or a Hubble patch after the emission.  The expression (\ref{Ent}) shows that these states 
 are entangled with outgoing quanta through the label $\ket{j}$.  \\
 
  The crucial point is that each $\ket{N',j}$ is also 
  self-entangled. That is, even if initial $\ket{N}$ was a tensor
  product state of various master and memory modes, 
 $ \ket{N_{a_1}} 
  \ket{N_{a_2}}... 
  \ket{n_{b_1}}  \ket{n_{b_2}}$, 
  in the state $\ket{N',j}$ they become entangled.  
  In other words,  the state evolves into a superposition,  
  \begin{equation} \label{vecE} 
  \ket{N'} = \sum_{a,b} c_{a_1,a_2,..b_1,b_2,...} \ket{N_{a_1}} 
  \ket{N_{a_2}}... 
  \ket{n_{b_1}}  \ket{n_{b_2}} \,,    
\end{equation} 
where $c_{a_1,a_2,..b_1,b_2,...}$ are coefficients.
We have deliberately ignored 
the label of the emitted species $j$, in order to focus 
on the internal entanglement. In addition, of course, the state is 
entangled with respect to the species index $j$. \\

Let the number of linearly-independent 
states participating in the superposition (\ref{vecE}) be $\tilde{n}_{\rm st}$. 
 We shall measure the level of the entanglement by the log of this number: 
\begin{equation} \label{S}
   S_{\rm ent} \equiv \ln(\tilde{n}_{\rm st}) 
 \end{equation}   
Obviously, $\tilde{n}_{\rm st}$ grows with each act of 
re-scattering and so does  $S_{\rm ent}$. 
The maximum is reached when the inner entanglement 
saturates the capacity of $N$-particle state.  
Obviously, $\tilde{n}_{\rm st}$ cannot exceed the total number of available 
micro-states, $n_{\rm st}$.  Thus, at its maximum,  the quantity 
$S_{\rm ent}$ equals to the micro-state entropy of the system
(\ref{SM}). \\

The shortest time for reaching the maximal entanglement,
\begin{equation} \label{EMAX}
   S_{\rm ent} = \frac{1}{\alpha}\,,
  \end{equation}   
 from 
an initial unentangled state, is bounded by unitarity. In an $N$-particle 
saturated system, with coupling 
constant $\alpha=1/N$ and frequencies $1/R$, this time is given by: 
 \begin{equation} \label{tmin}
   t_{\rm ent}  =   \frac{R}{\lambda}  = R \frac{N}{N_{\rm sp}} \,, 
 \end{equation}  
 where $\lambda$ is the species coupling (\ref{Lambda}). \\
   
 Intuitively, the above can be understood by noticing that 
 (\ref{tmin}) is the minimal time required for an order-one fraction of 
 the master modes to re-scatter.  $t_{\rm ent}$  is also a shortest time, 
 set by unitarity, required for a start of decoding the information stored in 
 a generic saturated system.  This will become much clearer 
 after considering the memory burden effect. \\

 As we see,  for de Sitter and black holes, the 
 time (\ref{tmin})  is equal to quantum break time
 $t_Q$.  It also matches the Page time.   We thus arrive to yet another physical meaning of this time-scale. \\

  The phenomenon of inner entanglement,  originates from 
 the multi-particle nature of the state $\ket{N}$. 
 This state hides complexity, which is not visible at initial times. 
 In order to resolve it, one needs to be sensitive to $1/N$ effects, which 
 are extremely suppressed for macroscopic values of $N$. 
 However, the complexity steadily grows in time.  
 The re-scattering  
   of constituents (which also leads to the emission of species $j$), simultaneously 
   generates the entanglement among them. 
   At the early times, the self-entanglement is small. However, 
  it grows 
   at the same rate (\ref{rate}) that governs the emission.
   The number of species can dramatically accelerate this growth.

 \subsection{Memory burden effect} 
 
 In the course of depletion, the gaps of the memory modes grow.  This happens, because the occupation number 
 of the master mode, $N$, diminishes.  Correspondingly, the system moves away from the saturation point (\ref{sat}). 
 Consequently, 
 the mechanism 
 behind the gaplessness of the memory modes is abolished.   \\
  
 This effect is clearly illustrated by 
 the equation  (\ref{Hgap1}).  The energy gap of a memory mode, 
  is non-zero away from criticality, 
 $N  <  1/\alpha$. 
When the master mode depletes, $N$ decreases and the
memory gaps grow.   
  Correspondingly, the energy cost of the memory pattern 
 (\ref{vec}) becomes higher and higher.  \\
 
For example, after reducing the occupation number of the master mode 
to half of (\ref{sat}), the gap (\ref{Hgap1}) would grow from zero 
to  $ \omega_b/4$.  
Correspondingly, the energy cost of the memory pattern (\ref{Hgap}) would become, 
 \begin{equation} \label{Epat}
 E_{\rm pattern} =  \bra{n_1,n_2,..} \hat{H} \ket{n_1,n_2,..} =
 \frac{1}{4}\sum_b  \omega_b n_b \,.
 \end{equation}
 This results into a back-reaction 
 force, called the ``memory burden" \cite{Dvali:2018xpy, Dvali:2018ytn,
 Dvali:2020wft}.  
This force resists to any departure of the system from criticality.  
In particular, it slows down the decay process. \\

   The memory burden exerts the maximum force, latest by
   the time the system looses about half of the quanta of its
   master mode.  This happens after the time of half decay 
   $t_{\rm half}$ (\ref{Half}).  As we already saw, this is also equal to the 
   time $t_{\rm ent}$, required for developing the inner entanglement.    
    Notice, since by $t_{\rm half}$ the memory modes acquire 
 non-negligible frequencies, the information pattern stored 
 in them starts to become readable.  
  Beyond this point, the system 
  cannot be described within any semi-classical approximation, even qualitatively.  The corresponding time-scale
  has a meaning of the quantum break-time $t_Q$ (\ref{QBT}). \\

 Some observational implications of the memory burden 
 effect for black holes has been studied in  \cite{Dvali:2020wft}. 
 In cosmology, the memory burden 
 can leave potentially-observable imprints from the inflationary epoch 
  \cite{Dvali:2018ytn, Prihadi:2020pqi}.
 These imprints, as well as other 
 quantum gravitational effects, are magnified by the number of species 
 $N_{\rm sp}$ \cite{Dvali:2020etd}.  The theories in which this number is large, 
 give better observational prospects for such effects.

\subsection{Physical meanings of $t_Q$} 

As we have seen, the three equal time scales,  
\begin{equation} \label{3t}
t_Q = t_{\rm half} = t_{\rm ent} =  \frac{R}{\lambda} \,,
\end{equation} 
represent different manifestations of one and the same corpuscular 
quantum clock. It is important that this time-scale 
stays finite in the species magnifier limit (\ref{MLimit}) of quantum gravity.  
This allows us to cleanly distill the quantum gravitational effects 
leading to it.  These effects are behind different physical meanings 
of the quantum clock.  We summarize them below.  

 \begin{itemize}

  \item 
 A time of half-decay of a saturated system.  
 
 \item  A quantum break-time: Time of full departure from the semi-classical regime. 
 
  \item 
  A time during which the memory burden reaches its maximum. 
 
  \item 
 A time after which an observer can start decoding the quantum information (memory pattern) stored in
  a saturated system.

  \item 
  A time during which a saturated system can reach the maximal 
  inner entanglement.

\end{itemize}

 \subsection{Comparing the Curves} 
    
As already explained, together with the memory burden effect, the inner entanglement 
is the main mechanism behind the Page's curve.  
 Both effects reach the maxima by the time $t_Q$. 
 Until this point, the resulting Page's curves for de Sitter and 
  black holes are very similar. The fundamental differences 
  start  beyond this point.  \\

   Unlike a black hole, de Sitter cannot exist 
  as a consistent state after $t_Q$. 
    The reason is the conflict 
  between the quantum state and the classical source 
  in form of a cosmological constant.  Because of this conflict, 
  the system must gracefully exit from the de Sitter phase, latest
  by the time 
  \begin{equation} \label{Exit}
   t_{E} \leqslant t_Q \,. 
 \end{equation}  
 Therefore, the continuation of the Page's curve beyond $t_Q$, depends on the mechanism
 of the graceful exit and the physics beyond it.
  For example,  the exit can be provided 
  by a time-dependent scalar field (inflaton).  \\
    
  The equation (\ref{Exit}) has profound cosmological implications. In particular, it puts an upper bound on the duration of inflation. This
excludes any sort of meta-stability.
Since $t_Q$ is an upper bound on duration of any given Hubble patch, 
it is impossible to satisfy in a sensible meta-stable state.   
 \\ 
  
  No such conflict arises for a black hole, which can continue its existence beyond $t_Q$. Of course, after this time,
due to maximal inner entanglement and memory burden, 
a black hole internally is no longer describable by any 
classical field configuration.  The semi-classical approximation 
breaks down fully.  In particular, the process of a decay, can no longer be 
described as Hawking's  thermal evaporation. \\

 Since the decay is not self-similar, we cannot even claim 
 that the black hole shrinks in size and heats up. 
 In fact, there are indications \cite{Dvali:2020wft}
that after $t_Q$ the decay of a black hole may slow down due to the ``memory burden" effect. 
  However, regardless of the slow-down, the entanglement, after reaching 
  the maximum, can only decrease. \\
   
  Let us now turn to the role of species in Page's curve. 
 This role is obvious. Since species shorten $t_Q$, they shorten
 (and sharpen) the entanglement curves for a de Sitter and  
 for a black hole.  
   
     \begin{figure}
 	\begin{center}
        \includegraphics[width=0.53\textwidth]{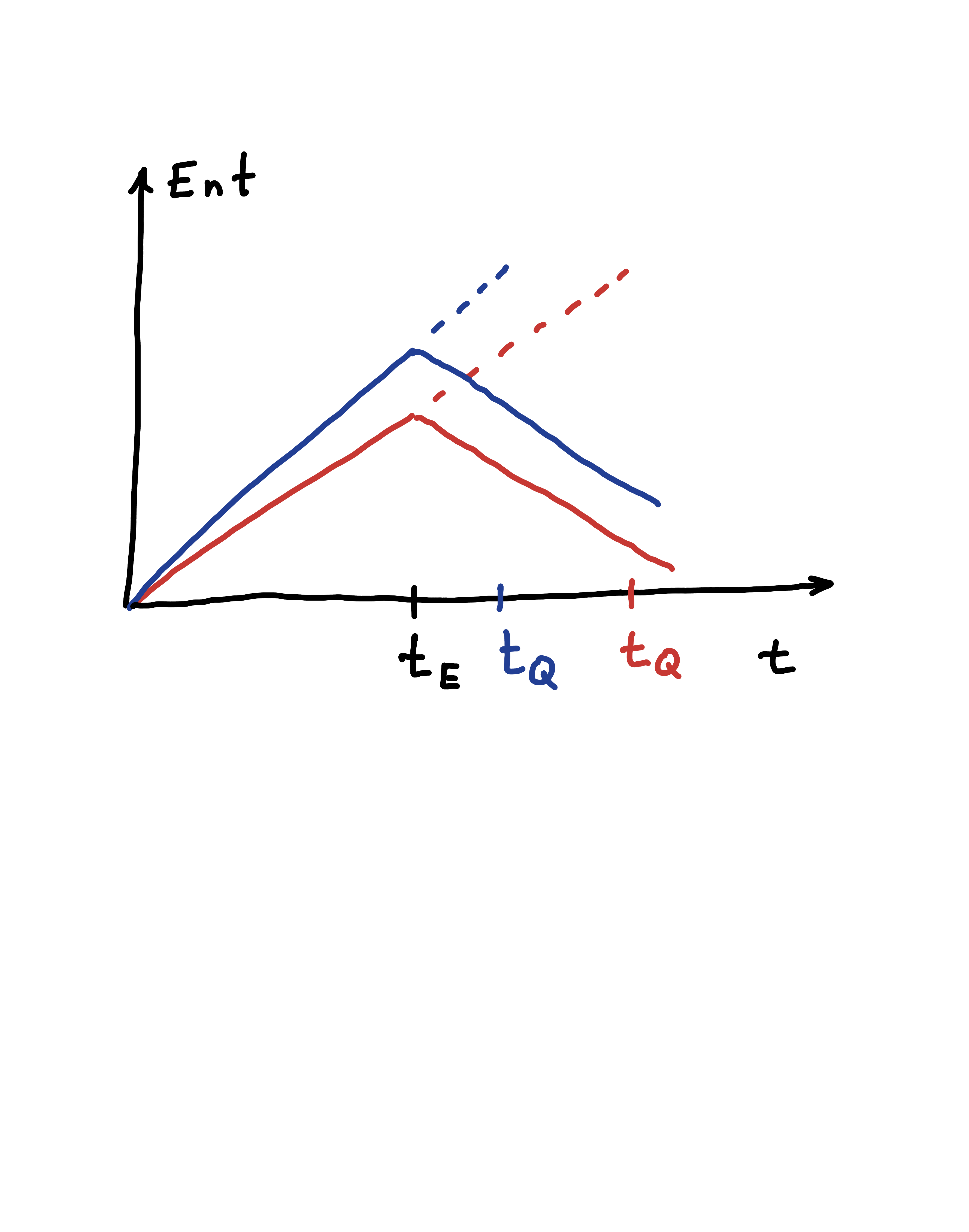}
 		\caption{A schematic description of Page's like curve for 
		de Sitter for different numbers of species. The blue curve 
		corresponds to a higher number  $N_{\rm sp}$ than the red one.  The corresponding quantum break-times $t_Q$ are denoted by the same colors  
as the curves. Unlike a black hole case, the curve cannot be extended 
beyond the quantum break-time $t_Q$. This time formally equals to a would-be Page's time but in de Sitter has a  fundamentally 
different effect.  
 The growth is terminated by the graceful exit that takes place at 
$t_{E} \leqslant t_Q$.  
	} 	
\label{2toN}
 	\end{center}
 \end{figure}

\section{String Theory} 

We now wish to discuss some implementations of the species magnifier 
limit (\ref{MLimit}) in string theory. 
We shall mostly follow \cite{Dvali:2020etd} in which 
the needed expression for string theoretic 
quantum break-time has been derived. 
A complementary derivation of quantum break-time 
(\ref{QBT}) in string theory context, was 
also given in \cite{Blumenhagen:2020doa}. \\

The regime (\ref{MLimit}) shows transparently, how the string theory 
responds to a deformation towards a de Sitter type state.   For producing such a deformation, we shall use the standard 
uplifting of the energy via the 
tension of (anti)$D$-branes \cite{Dvali:1998pa}. \\

For definiteness, we shall work in type {\it II}B in ten dimensions, with string coupling $g_s$ and 
the string scale $M_s$.  
As in  \cite{Dvali:2020etd}, we shall consider the background 
with number of coincident $D_9-\bar{D}_9$-brane pairs. 
The world-volume (open string) physics of this system is 
rather well-understood (see, \cite{Srednicki:1998mq} and 
subsequent papers). Our task is to highlight how the system 
responds to ``de Sitterization".  One response is
a finite quantum break-time, which was already derived in  \cite{Dvali:2020etd}.  Here, we wish to show that this time-scale carries a
string theoretic information about the Gibbons-Hawking 
entropy. 
  \\ 

In string construction, the role of $M_*$ is played by $M_s$. 
 The roles of the low energy species are 
 played by the zero modes of the open strings.  With $n$ $D_9$-branes 
and $n$ anti-$D_9$-branes  piled-up on top of each other, there exist $N_{\rm sp} \sim n^2$ 
light  Chan-Paton species. They come from the zero modes of the open strings and transform under the $U(n)\times U(n)$ 
gauge symmetry.  
 \\

Of course, such a configuration is unstable since $D$-branes and 
anti-$D$-branes can annihilate each other. 
This instability is manifested by an open string tachyon.
The $D-\bar{D}$ annihilation can be described as tachyon 
condensation \cite{Sen:1999mg}.
During this condensation the gauge symmetry is partially Higgsed and 
the Chan-Paton species become massive. 
In the original ``brane inflation" scenario  \cite{Dvali:1998pa}, 
the tachyon condensation was used as the graceful exit mechanism
from the inflationary phase. \\

In the present discussion, the main interest is the un-condenced
phase, in which tachyon is placed on top of the ``hill".
 Classically, such a phase can produce a de Sitter like
 metric  with the curvature radius, 
 \begin{equation} \label{stringR} 
   R =  \frac{1}{\sqrt{ng_s} M_s} \,.
    \end{equation}  
 Viewing this as a coherent state of gravitons, we can estimate 
 the number $N$ of the constituent master modes. 
 Since each constituent brings roughly the energy 
 $1/R$, we can determine $N$ by matching $N/R$ with the energy of the Hubble patch. 
 This gives, 
  \begin{equation} \label{stringN} 
   N  =  \frac{1}{g_s^2(ng_s)^4} \,.
    \end{equation}  
  It is easy to see that this quantity equals to a Gibbons-Hawking entropy 
  of a would-be ten dimensional de Sitter with the radius 
  (\ref{stringR}):
  \begin{equation} \label{stringSGH} 
   S_{GH}  =  (R M_{10})^8 = \frac{1}{g_s^2(ng_s)^4} \,,
    \end{equation}   
   where $M_{10}$ denotes the ten-dimensional Planck mass,
   related to $M_s$ and $g_s$ through, 
   \begin{equation} \label{MsMp} 
   M_{10}^8  = \frac{M_s^8}{g_s^2} \,.
    \end{equation}

   On the other hand, the ten-dimensional quantum coupling of gravitons of wavelength $R$ (\ref{stringR}), is given by, 
  \begin{equation} \label{stringAplha} 
   \alpha  = \frac{1}{(R M_{10})^8} =  g_s^2(ng_s)^4 \,.
    \end{equation}  
The equations (\ref{stringN}), (\ref{stringSGH}) and (\ref{stringAplha})
confirm that a (would-be) de Sitter obtained in this construction 
is a saturated state
(\ref{sat}). \\

 At the same time, the species coupling 
(\ref{Lambda}) is equal to 
 \begin{equation} \label{stringLambda} 
   \lambda  = \alpha n^2 = (ng_s)^6 \,.
    \end{equation}  
    
 The corresponding quantum break-time, 
according to (\ref{QBT}), is given by 
 \begin{equation} \label{QstrD} 
   t_Q = \frac{1}{M_s}\frac{1}{(n g_s)^{13/2}} = 
   \frac{R}{(ng_s)^6} = \frac{R}{\lambda} \,.
    \end{equation}
 This expression was already derived in \cite{Dvali:2020etd}.  \\
   
  We shall now take the species magnifier limit (\ref{MLimit}). 
 In the present case, this limit reads:        
   \begin{eqnarray}
  \label{stringMLimit}  
 && n  \rightarrow \infty\,, ~g_s \rightarrow  0\,, ~ \lambda = (ng_s)^6 = {\rm finite} \,,\\ \nonumber 
 && M_s = {\rm finite}\,,~ R  = {\rm finite}\,. 
 \end{eqnarray}
In this limit, all stringy processes that are not accompanied by proper  
powers of $N$ or $n$, vanish. 
  Notice, the quantum break-time (\ref{QstrD}) remains finite. \\
  
 For $ng_s \ll 1$, the quantum break-time
 (\ref{QstrD}) is much longer than the
 $D$-brane instability time due to tachyon, which is of order $1/M_s$.
Thus, unless additional stabilizing measures are taken, the 
tachyon condenses prior to quantum breaking.  That is, the 
tachyon condensation provides a graceful exit that saves the 
would-be de Sitter from quantum breaking. \\

What shall happen if we try to stabilize the  tachyon?
This can be attempted in the following way. 
Notice that the tachyon coupling to the Chan-Paton species 
generates an additional contribution to the mass-square of the tachyon 
field. This is due to their production 
via Gibbons-Hawking mechanism.
The effect can be approximated as coming from a 
thermal bath of temperature $T_{GH} =1/R$. 
This generates an effective thermal 
mass for the tachyon,  
\begin{equation} \label{DeltaM} 
  \Delta m^2_{tachyon} = 
  \frac{(ng)^2}{R^2} =  (ng_s)^3 M_s^2 \,.
    \end{equation}     
 This contribution is positive and could potentially balance the zero-temperature negative 
 mass-square of the tachyon, provided 
 \begin{equation} \label{CPLimit} 
  (ng_s) \sim 1 \,.
    \end{equation}  
  In this case, the stack of branes could be stabilized.\\
  
  This stabilization can be viewed as a restoration 
  of $U(n)\times U(n)$ symmetry at high temperature. 
  Such a stabilization of $D$-branes of top of each other by thermal effects has been considered previously (see second reference in  \cite{Dvali:1998pa}). The difference in the present case is that the 
  stabilization would be ``self-imposed" in the sense that it would come from the Gibbons-Hawking temperature 
  created by the brane configuration. That is, branes create de Sitter, which creates a thermal density of open string modes. These   
  thermal corrections, in turn,  
  stabilize the tachyon. This sounds too good to be true. 
  And indeed, there is a caveat. \\

  Notice \cite{Dvali:2010vm}, (\ref{CPLimit}) represents the point for which the number of Chan-Paton species 
  saturates the ten-dimensional version
    of the species bound (\ref{Mstar}),
  \begin{equation} \label{M10} 
  M_* =  \frac{M_{10}}{N_{\rm sp}^{1/8}} \,.
    \end{equation}  
    Taking into account that $M_* = M_s$ and using the relation 
  (\ref{MsMp}), we translate (\ref{M10}) as the limiting relation (\ref{CPLimit}) between $n$ and $g_s$. 
    \\

 Thinking of the relation (\ref{CPLimit}) as the limit,
  we notice a rather interesting tendency.   
  This is about entropy of a would-be de Sitter state. 
 Due to the existence 
   of the light Chan-Paton species, we have a  
   non-gravitational contribution to the micro-state entropy. 
   This contribution, roughly, scales as their number \footnote{
  More precisely, around the saturation point (see below) the 
  entropy scales as $S_{CP} \simeq \ln\begin{pmatrix}
    n^2  + \frac{1}{g_s^2}   \\
      n^2  
\end{pmatrix}$.}, 
   \begin{equation} \label{CPentropy}
   S_{CP} = n^2 \,.
  \end{equation}    
  We observe that, in the limit (\ref{CPLimit}), the 
  entropy of the open string Chan-Paton species (\ref{CPentropy}), 
matches the Gibbons-Hawking 
  entropy (\ref{stringSGH}) of a would-be de Sitter state: 
    \begin{equation} \label{CPGH}
   S_{CP} = S_{GH} = \frac{1}{g_s^2} \,.
  \end{equation}     
  However, at this point, the two critical effects take place in 
 UV and  IR theories.   \\ 
   
 First, the curvature
 becomes of order the string scale.  This is the warning sign from 
 the UV theory.  We expect that, beyond this 
 threshold, any notion of the 
 de Sitter geometry is lost.    
   This expectation is fully supported by the knowledge \cite{Bowick:1992qu}
  that 
  the Hagedorn effects \cite{Hagedorn:1965st}
 set in above the string temperatures. 
 The fundamental stringy effects, such as Atick-Witten phase transition \cite{Atick:1988si},
 must be taken into account.  The discussions of aspects of 
 thermal phase can be found in    
 \cite{Alvarez:1986sj}, \cite{Dienes:2012dc},  \cite{Blumenhagen:2020doa} (and references therein). 
  Whatever the theory is in this phase, it is not de Sitter. \\

 It is remarkable, how the low energy theory accounts for this breakdown.  First, this is signalled by the fact that the quantum break-time
 (\ref{QstrD}) becomes of order
  the Hubble radius, which is also given by the string length,
   $t_Q \sim 1/M_s$.  \\

 The second indication from the low energy theory is the 
 behaviour of the  
 scattering amplitudes.  Namely, exactly at this point, the processes of
  graviton-graviton scattering into Chan-Paton species,
  saturate unitarity bound (\ref{sat}). The example is given by a process 
   in which two gravitons scatter into many Chan-Paton quanta
  with total occupation number $N=1/\alpha$.  \\
  
 This process falls in the category of generic 
 $2 \rightarrow N$ processes of the type (\ref{2inN}). 
 As already discussed in section \ref{SLimit}, they saturate unitarity 
 when the entropy of species reaches a critical value (\ref{sat}) 
 \cite{Dvali:2020wqi}. 
 Near the saturation point the cross-section is given by 
 (\ref{crossA}). In the present case, $\alpha$ is given by (\ref{stringAplha}), whereas the entropy $S$
 is given by (\ref{CPentropy}). 
  Therefore, around saturation,  the cross section can be written as, 
  \begin{equation} \label{cross2}
   \sigma  = {\rm e}^{-n^2 \left (\frac{1}{(ng_s)^6} - \frac{S_{CP}}{n^2} \right )} 
 =  {\rm e}^{-n^2\left (\frac{1}{\lambda} - \frac{S_{CP}}{n^2}\right )} \,.
  \end{equation}    
Taking into account (\ref{CPentropy}), the unitarity is saturated for $\lambda \sim 1$.  Or equivalently, at the saturation of unitarity 
by the cross section (\ref{cross2}), 
 the open string 
Chan-Paton entropy matches the Gibbons-Hawking one
(\ref{CPGH}). \\

 Through the above phenomenon,  the theory is telling us something profound. 
   In the light of open-closed correspondence, which is intrinsic 
  to string theory, we would expect that the open string sector would 
  carry information about the Gibbons-Hawking entropy 
  of de Sitter vacuum, if such were to exist.
  And, indeed, we see that the theory does contain such a resource, through the entropy of the open string
  modes (\ref{CPentropy}). However, this entropy matches the 
  would be Gibbons-Hawking entropy, exactly at the point
  where unitarity is saturated. Simultaneously, 
  the curvature (\ref{stringR})  becomes stringy and so does  the quantum 
break-time (\ref{QstrD}). As a result, the chance of getting a
stable de Sitter state is abolished.  
 This fully matches the conclusion from the previous arguments  
 \cite{Dvali:2020etd} that de Sitter vacuum 
 is incompatible with non-trivial closed string $S$-matrix.  \\
 
 Notice, none of the above obstructions appear when accounting for 
 Bekenstein-Hawking entropy of an extremal supersymmetric 
 black hole by string degrees of freedom, as it is known from
 Strominger-Vafa construction \cite{Strominger:1996sh}.
This difference, is also connected with the fact that de Sitter cannot preserve any supersymmetry and, 
unlike an extremal black hole,  exhibits a finite quantum break-time.  \\

Similarly, no quantum breaking is evident to take place (at least in supersymmetric limit)
by an anti de Sitter, physics of which is accounted by 
CFT \cite{Maldacena:1997re, Gubser:1998bc, Witten:1998qj}. 
 We shall leave open the question of possible connection 
 between this physics and the ideas of saturation and corpuscular structure. Some preliminary thoughts can be found in 
\cite{Dvali:2011aa, Dvali:2013eja}.

 \section{Some Observational Implications}

 In standard slow-roll picture, the inflaton plays 
 a double role.  First, it provides a classical clock that enables 
 inflation to end.  
  Secondly, the quantum fluctuations of this 
 clock provide density perturbations. 
 The quantum corpuscular effects, enhanced by species, interfere
 with both mechanisms
 of the inflaton clock. 
 \\

 \subsection{Enhancement of Gibbons-Hawking Perturbations} 
 
 First, the density fluctuations 
from  Gibbons-Hawking production of species, is enhanced by their number,
 \begin{equation}\label{Dsp}
 \delta_{\rm sp}^2 = \frac{H^2}{M_P^2} N_{\rm sp} \,.  
 \end{equation} 
 No such enhancement is exhibited by inflaton fluctuations, 
  \begin{equation}\label{Dinf}
 \delta_{\rm inf}^2 = \frac{H^2}{\epsilon M_P^2} \,, 
 \end{equation} 
where  $\epsilon \equiv (V'M_P/V)^2$ is the standard inflationary slow-roll parameter.  \\

  Using (\ref{Mstar}), it is instructive to rewrite the
 above expressions as functions of $M_*$,  
 \begin{equation}\label{DinM}
 \delta_{\rm sp}^2 = \frac{H^2}{M_*^2} \,,~
  \delta_{\rm inf}^2 = \frac{1}{\epsilon N_{\rm sp}} 
 \frac{H^2}{M_*^2} \,.  
 \end{equation} 
We see that in the limit (\ref{MLimit}), which implies 
$\frac{1}{\epsilon N_{\rm sp}}  \rightarrow  \infty $, 
the iflaton perturbations  vanish. 
  \\

We wish to comment that, in case of a large 
number of light spin-2 species, the enhancement could potentially 
be experienced by the tensor mode \cite{Starobinsky:1979ty}.

\subsection{Quantum Clock} 

 As we have seen,  the species speed up a different clock in 
 de Sitter. 
 This is an internal quantum clock originating from the corpuscular
 effects such as depletion, self-entanglement and memory burden. 
This quantum clock advances regardless of the classical one 
 of a slowly-rolling inflaton field.
If the number  $N_{\rm sp}$ is above a certain critical 
 value, the quantum clock elapses faster than 
the classical one. This puts an upper bound on $N_{\rm sp}$.  \\

  In general, if the number of species is large, 
the effect of the 
quantum clock cannot be ignored.  
 In particular,  the tilt in the spectral index 
 will include contributions from two different sources: The standard 
 classical tilt,  due to change of $H$, and the quantum one 
 which we shall now quantify.  \\
  
 The quantum clock gives a 
 departure from a (would-be) scale-invariant spectrum
 of perturbations.  This departure is unrelated to a classical slow-roll 
 clock and is due to depletion of the de Sitter constituents. 
 The corresponding 
 tilt in the spectral index can be estimated as, 
  \begin{equation}\label{tilt} 
  (n_s - 1)_{\rm Q} =  - \frac{N_{sp}H^2}{M_P^2} \,.
    \end{equation} 
  This expression for $N_{\rm sp} =1$ was derived in 
  \cite{Dvali:2013eja}.  The effect is enhanced by the number of species
 \cite{Dvali:2020etd}.  \\

 We shall try to illuminate this phenomenon from various angles.  
For this, we represent (\ref{tilt}) in 
three equivalent ways, 
 \begin{eqnarray} \label{T}
(n_s - 1)_{\rm Q} & = & - 1/(t_Q H) \\ \nonumber
 & = & - \frac{H^2 }{M_*^2} \\  \nonumber
 &=&  \lambda \,. 
\end{eqnarray} 
An each form, highlights a different physical meaning of the same 
phenomenon.  We shall review them separately. \\

 The first equality in (\ref{T}) shows that the 
tilt is the ratio of Hubble-time $1/H$ to quantum break-time $t_Q$.  
This makes sense, since, according to (\ref{rateC}), 
after time $t_Q$ the particle-creation rate drops by order one. 
The effect of the number of species on the tilt is 
via shortening $t_Q$, as it is clear from (\ref{QBT}).  
 \\
 
The second equality in (\ref{T}), tells us that the tilt   
is the square of the ratio of Hubble  $H$ to gravity cutoff
$M_*$.  According 
to (\ref{Mstar}),  the job of the species is to lower $M_*$. 
Correspondingly, they increase the tilt. \\ 

 Finally, the last equality in (\ref{T}) tells us that the  
 tilt is equal to the species coupling $\lambda$. 
 Again, the role of the species in increasing tilt is transparent 
 through  their contribution  into  $\lambda$, as 
 described by (\ref{Lambda}).   \\
 
 Note, the information pattern  carried by de Sitter is expected 
to contribute into the tilt through the memory burden effect, 
 \cite{Dvali:2018xpy, Dvali:2018ytn, Dvali:2020wft, Prihadi:2020pqi}. 
 Some alternative descriptions can be found in \cite{Gomez:2021yhd}. \\

\subsection{Phenomenological Bounds}

 The large numbers of particle species are included in  
 many extension of the Standard Model. 
 The extreme case is provided by the scenario 
  \cite{Dvali:2007hz} which incorporates         
    \begin{equation} \label{32}
    N_{\rm sp} \sim 10^{32} \,.
   \end{equation} 
This number is motivated by the Hierarchy Problem, an outstanding puzzles of the Standard Model. The  essence of it
is that the mass term of the Higgs scalar 
is quadratically sensitive to the UV-cutoff of the theory.  
Ordinarily, the role of the cutoff would be played by $M_P$. 
Hence, having the Higgs mass around $M_P$, would raise no concern. 
However, the Higgs is much lighter than $M_P$. 
This is the source of the puzzle. \\

 The idea of  \cite{Dvali:2007hz} is to use
 the fact that species lower the cutoff 
 (\ref{Mstar}). 
 In a theory with $N_{\rm sp}$ given by 
 (\ref{32}), the cutoff $M_*$ is pushed 
down to  its current experimental bound of few TeV.  
In this case, the Hierarchy Problem gets effectively nullified. \\

Ordinarily, the inflationary cosmology is assumed to be 
associated with very high energy scales.  In this light, a theory 
with low cutoff raises concerns because the Hubble parameter 
has to be small. The perception is that 
such a theory has a difficulty of generating density perturbations. 
 Again, this concern is coming from an implicit assumption 
 that density perturbations originate from the fluctuations of
 a single slow-rolling scalar field.  
However,  the same physics that pushed the Hubble down, 
also provides  an alternative 
mechanism of enhancing the quantum gravitational particle-creation 
by the number of species.  This needs to be taken into account
in inflationary model building. \\ 
 
 Since (\ref{32}) represents an absolute phenomenological upper bound on  the number of particle species, it is useful to give some examples  
 of enhancement due to this number.  
  For instance, for (\ref{32}),  
the observable quantum gravitational 
imprints can be generated by inflation with the  Hubble parameter 
in a range of nuclear energies, or so. \\

 As another example, 
a primordial black hole of $10^{-3}$ earth mass, would currently  
approach its quantum break-time and be at the peak of its Page's
curve.  Despite being macroscopic, such a black hole 
would represent a fully quantum object, with a maximal inner 
entanglement.  Also a memory burden effect would be maximal. 
A scattering of radiation at such 
a black hole, would in principle trigger an induced 
radiation coming from the memory modes. 
Similar considerations apply to lighter black holes. \\

The observational prospects, of course, depend on the number density 
of such black holes.  As the preliminary studies indicate
\cite{Dvali:2020wft}, the 
memory burden effect slows their decay after $t_Q$.  
As suggested, this effect can potentially enlarge the allowed mass window of primordial black holes as dark matter 
\cite{Hawking:1971ei, Carr:1974nx, Carr:2016drx}
towards the lighter end.

 \section{Outlook: Background Matters}
 
 In the present paper we have discussed quantum gravity in 
 species limit (\ref{SP1}). This allows to cleanly isolate certain collective 
 and non-perturbative effects that otherwise would be shadowed by irrelevant corrections. 
The example is a quantum break-time (\ref{QQQ}), 
exhibited by saturated systems such as de Sitter and black holes
\cite{Dvali:2013eja, Dvali:2014gua, Dvali:2017eba, Dvali:2020etd}. 
 The species limit (\ref{SP1}) allows to keep $t_Q$ finite, 
 despite the fact that the perturbative gravity decouples.  \\
 
   Working in the species limit, allows to visualize the time-scales that control the inner entanglement and the back-reaction 
from the information pattern (memory burden).  These are the corpuscular mechanisms that lead the system to quantum breaking.   
    \\

   We extracted the quantum break-time  
 in string theoretic example in which the species are 
 counted by the Chan-Paton factors.  
  This example shows how profoundly the string theory resists 
 to formation of a would-be de Sitter vacuum.  
 The synchronized warning signs are given by 
 low and high energy theories. \\
 
  The quantum break-time, computed in low energy theory of zero-modes, becomes of order the string length
 when the high energy theory enters the Hagedorn phase.   
  At the same time, the entropy of the open string modes matches the Gibbons-Hawking 
  entropy.  However, this matching takes place 
  at the saturation point of the ten-dimensional 
  species bound (\ref{M10}).  This is accompanied by the saturation of 
 the unitarity bound (\ref{S111}) by the entropy of the low energy species.  \\
   
 The above observations, reinforce the earlier conclusions
 \cite{Dvali:2013eja, Dvali:2014gua, Dvali:2017eba, Dvali:2020etd}
that the finite quantum break-time is the manifestation of inconsistency 
of the de Sitter state. 
The same message is carried by more recent 
conjectures \cite{Palti:2019pca}. 
 \\

 Next, we wish to put the concept of the exclusion of de Sitter
 by the $S$-matrix, in a broader perspective
 of background dependence of quantum gravity. 
 The first point is to take into account the double-scaling 
argument  \cite{Dvali:2020etd},
 which shows that ``vacuumization" of de Sitter is linked with trivializing the graviton (closed string) $S$-matrix. Ordinarily, we assume that an 
approximate $S$-matrix treatment should be valid at short distances. 
This is true,  as long as the short-distance limit commutes with the formulation of the theory.  \\

That is, it is OK to perform an approximate computation, 
after the theory has been formulated properly.  However, 
there is no {\it a priory} justification for doing the same, within
a theory that is approximately formulated. 
 The last concept not always makes sense. 
 A good example of such an inconsistency, would be a gauge theory that is  {\it approximately
anomalous}. In fact, the connection between quantum break-time 
of de Sitter and anomaly was already made in \cite{Dvali:2020etd}, by  
drawing a parallel with anomalous $1/N$-effects of Witten-Veneziano mechanism \cite{Witten:1979vv}. \\

  The second lesson is that, not every background 
  of classical General Relativity, must find its vacuum realization within  
  quantum gravity/string theory.  Quantum gravity is more fundamental 
  and more restrictive.  It is therefore natural that 
  only certain backgrounds of the classical theory can be given a tittle of a consistent vacuum. Many backgrounds, such as de Sitter, 
  that cannot fit in the vacuum category, must be viewed as 
  excited states constructed on top of a valid $S$-matrix vacuum. 
  They will be subject to severe constraints such as the
  quantum break-time \cite{Dvali:2013eja}. \\
  
   The non-acceptance  of all possible classical
  backgrounds as vacua, must be taken as an advantage 
 of the theory. 
  Needless to say,  by eliminating de Sitter vacua, the quantum gravity/string theory 
  nullifies an outstanding cosmological puzzle.   \\ 
  
  This suggests that the requirement of a background-independent formulation 
  of quantum gravity may not necessarily be the only viable strategy.
  Rather, an impossibility of formulation of a theory on a given background, can be viewed as a selection tool that increases the 
 restrictive power of the theory.  \\

 {\bf Acknowledgements} \\
 
 We thank Goran Senjanovi\'c for valuable discussions and comments.
We thank  Alexey Starobinsky for conversation on possible enhancement 
of the tensor mode. 
This work was supported in part by the Humboldt Foundation under Humboldt Professorship Award, by the Deutsche Forschungsgemeinschaft (DFG, German Research Foundation) under Germany's Excellence Strategy - EXC-2111 - 390814868,
and Germany's Excellence Strategy  under Excellence Cluster Origins.

\end{document}